\documentclass[review,number,sort&compress]{elsarticle}

\usepackage{graphicx}
\usepackage{dcolumn}
\usepackage{bm}
\usepackage{relsize}
\usepackage{mdwlist}
\usepackage{amsmath}
\usepackage{amssymb}
\usepackage{color}
\usepackage{soul}
\usepackage{pifont}
\usepackage{hyperref}
\usepackage[all]{hypcap}

\def\cesrta{{C{\smaller[2]ESR}TA}}
\def\degree{{^\circ}}

\begin{document}


\title{Measurements of Electron Cloud Growth and Mitigation in Dipole, Quadrupole, and Wiggler Magnets}

\author{J.~R.~Calvey}
\author{W.~Hartung}
\author{Y.~Li}
\author{J.~A.~Livezey\fnref{berk}}
\fntext[berk]{Present address: Department of Physics, University of California, Berkeley, CA, USA, 94720} 
\author{J.~Makita\fnref{odu}}
\fntext[odu]{Present address: Department of Physics, Old Dominion University, Norfolk, VA, USA, 23529}
\author{M.~A.~Palmer\fnref{fnal}}
\fntext[fnal]{Present address: Fermi National Accelerator Laboratory, P.O. Box 500, MS 221, Batavia, IL, USA, 60510} 
\author{D.~Rubin}
\address{Cornell Laboratory for Accelerator-based Sciences and
  Education, Cornell University, Ithaca, New York, USA, 14853}

\date{\today}

\begin{abstract}
Retarding field analyzers (RFAs), which provide a localized measurement of the electron cloud, have been installed throughout the Cornell Electron Storage Ring (CESR), in different magnetic field environments.  This paper describes the RFA designs developed for dipole, quadrupole, and wiggler field regions, and provides an overview of measurements made in each environment.  The effectiveness of electron cloud mitigations, including coatings, grooves, and clearing electrodes, are assessed with the RFA measurements.

\end{abstract}

\maketitle


\section{\label{sec:intro} Introduction}

The electron cloud effect, in which a high density of low energy electrons builds up inside a vacuum chamber~\cite{ECLOUD12:Tue1815}, has caused operational difficulties at a number of past and present accelerators~\cite{PhysRevSTAB.7.124801}, and is expected to be a limiting factor in next generation machines~\cite{ILCREP2007:001}.  As part of the CESR Test Accelerator (\cesrta) program at Cornell~\cite{CLNS:12:2084}, the Cornell Electron Storage Ring (CESR) was instrumented with a number of electron cloud detectors~\cite{NIMA760:86to97,PRSTAB17:061001,NIMA749:42to46,ARXIV:1407.0772,NIMA754:28to35}, including retarding field analyzers (RFAs)~\cite{NIMA453:507to513}.  Previous papers have discussed the design of RFAs for measurements in magnetic field free regions~\cite{NIMA760:86to97}; and described the use of simulations to quantify the electron emission properties of different cloud mitigating coatings tested in these regions~\cite{PRSTAB17:061001}.  This paper will summarize results obtained from RFAs located in dipoles, quadrupoles, and wigglers.

Electron cloud buildup has been measured in dipole and wiggler fields at SLAC~\cite{NIMA621:33to38}, KEK~\cite{NIMA598:372to378,Suetsugu2009449}, INFN~\cite{PhysRevLett.110.124801}, and CERN~\cite{PhysRevSTAB.11.094401,ECLOUD02:17to28}; and in quadrupole fields at LANL~\cite{PhysRevSTAB.11.010101} and LBNL~\cite{PhysRevLett.97.054801}.  At CESR we have worked to replicate and expand on these measurements, with the goal of providing specific recommendations for cloud mitigations in the International Linear Collider~\cite{ILCREP2007:001} damping ring (ILC DR), as well as deepening our understanding of cloud dynamics in magnetic fields.

\subsection{Retarding Field Analyzers}

A retarding field analyzer consists of three main components~\cite{NIMA453:507to513}: small holes drilled in the beam pipe to allow electrons to enter the device; a ``retarding grid," to which a voltage can be applied, rejecting electrons with less than a certain energy; and a positively biased collector, to capture any electrons which make it past the grid.  If space permits, additional (grounded) grids can be added to allow for a more ideal retarding field.  In addition, the collectors of most RFAs used at \cesrta~are segmented transversely to allow characterization of the spatial structure of the cloud build-up.  Thus a single RFA measurement provides information on the local cloud density, energy, and transverse distribution.

\subsection{Experimental Sections}

There are four main electron cloud experimental sections of CESR that contain RFAs in magnetic fields.  These are:


\begin{itemize}
    \item An detector inside a CESR dipole in the B12W arc section.
    \item A chicane of four dipole magnets, each containing an RFA, located in the L3 straight.
    \item An instrumented quadrupole (Q48W), also in L3.
    \item Three superconducting wigglers, each containing 3 RFAs, in the L0 straight.
\end{itemize}

\begin{figure}
	\centering
	\includegraphics[width=0.45\textwidth]{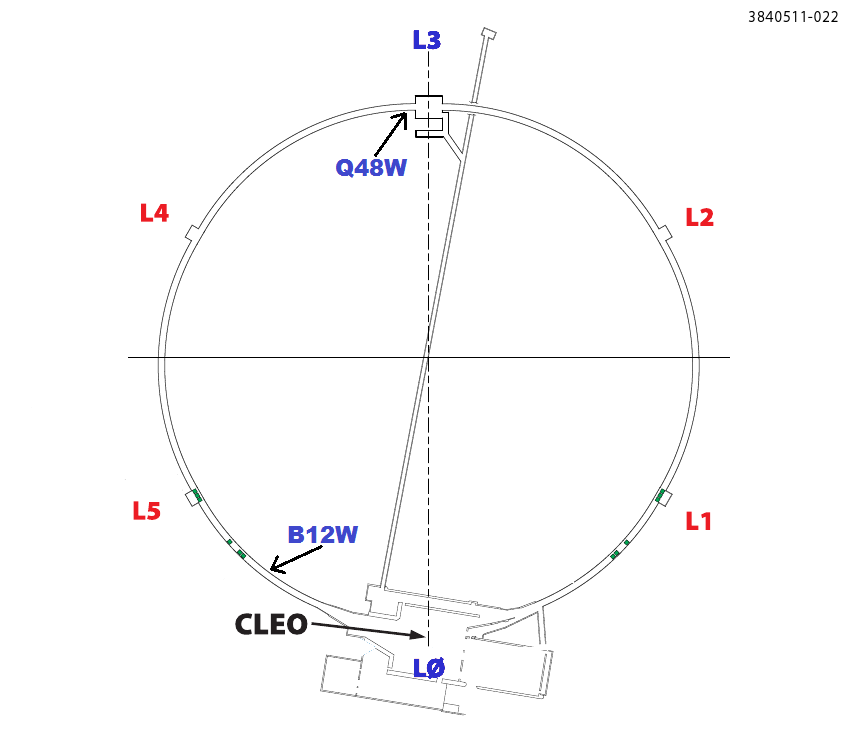}     
	\caption[{\cesrta} Vacuum System]{Present layout of the CESR ring.  Electron cloud detectors in magnetic fields have been installed in L0, B12W, Q48W, and L3.\label{fig:cesr_configuration}}
\end{figure}

The properties of each instrumented chamber are listed in Table~\ref{tab:rfa_master_list}.  Fig.~\ref{fig:cesr_configuration} shows the locations of these experimental sections in the CESR ring; more details on each location are given below.

\begin{table}
  \centering
  \footnotesize
    \caption[List of RFA locations]{List of RFA locations.  The ``Chamber" column refers to the base material; some locations have tested one or more mitigations.  The vacuum chambers at all locations are 5~cm in height by 9~cm in width, with the exception of the circular chambers, which are 4.5~cm in radius.}\label{tab:rfa_master_list}
  \begin{tabular}{cccc}
  \hline \hline
  Location  &   Magnet   &  Field Strength & Chamber  \\
  \hline
  L0    &   Wiggler         &   1.9 T (peak)                            &   Rectangular Cu \\
  12W   &   CESR Dipole     &   0.079~T (2.1~GeV) - 0.201~T (5.3~GeV)     &   Elliptical Al \\
  L3    &   Quadrupole      &   3.7 T/m (2.1~GeV) - 7.4 T/m (5.3~GeV)   &   Circular Al   \\
  L3    &   Chicane Dipole  &   0 - 0.1460~T                             &   Circular Al   \\
  \hline \hline
  \end{tabular}
\end{table}

\subsubsection{\label{ssec:dip_sec} B12W and L3 Dipole RFAs}

In the presence of a dipole magnetic field, an electron will undergo helical motion, spiralling around the field lines.  For a standard dipole magnet in an accelerator (with strength $\sim$ 1~kG), a typical cloud electron (with energy $\sim$ 10 - 100~eV) will have a cyclotron radius on the order of a few hundred $\mu$m.  In other words, the motion of the electron will be approximately one dimensional, along the direction of the dipole field.  This ``pinning" of the motion to the field lines results in an electron cloud buildup that is both qualitatively and quantitatively different from the field free case.

To study cloud buildup in such an environment, a thin RFA was designed to fit inside a CESR dipole magnet (Section~\ref{sssec:cesr_dip}), and installed in the 12W section of CESR.  The magnetic field at this location depends on the beam energy: 790~G at 2.1~GeV, 1520~G at 4~GeV, and 2010~G at 5.3~GeV.  The chamber is made of uncoated (6063) aluminum.

\begin{figure}
	\centering
    \begin{tabular}{c}
    \includegraphics[width=0.7\textwidth]{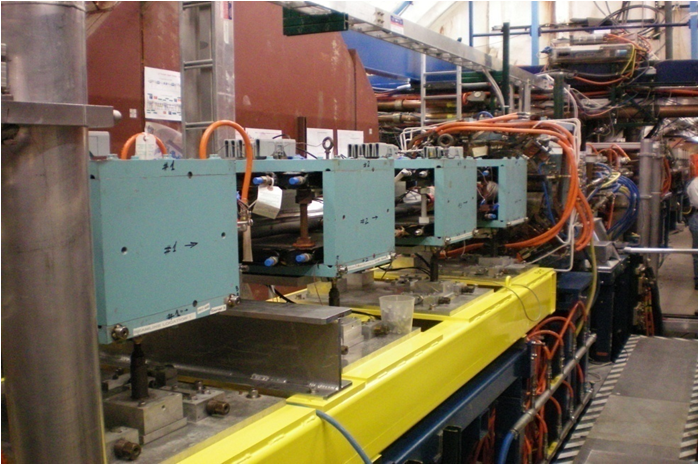} \\
    \end{tabular}
	\caption{PEP-II 4-dipole magnet chicane and RFA-equipped vacuum chambers. \label{fig:SLAC_Chicane}}	
\end{figure}

Additionally, a chicane of four magnets designed at SLAC~\cite{NIMA621:33to38} was installed in the L3 straight (Section~\ref{sssec:chic_dip}).  The field of these magnets can be varied over the range of 0 to 1.46~kG, which allowed for the study of the effect of dipole field strength on cloud dynamics, without affecting the trajectory of stored beams.    A photo of the chicane is shown in Fig.~\ref{fig:SLAC_Chicane}.


\subsubsection{Q48W Quadrupole RFA}

As in the dipole case, cloud electrons in a quadrupole will be constrained to spiral along the field lines.  However, the non-uniform character of the quadrupole field can lead to unexpected behavior.  Of particular interest is long term cloud trapping.  To investigate  these effects, a thin RFA was built to fit inside the Q48W quadrupole in CESR (Section~\ref{ssec:quad_inst}).

\subsubsection{\label{sssec:l0} L0 Wiggler RFAs}

Wigglers are an important component of next generation lepton colliders, as they greatly increase the radiation damping of the beam~\cite{ILCREP2007:001}.  The high rate of photon production in wigglers, combined with the complex three dimensional nature of their magnetic fields, makes electron cloud buildup inside them a serious concern~\cite{PRSTAB14:041003}.

\begin{figure}
	\centering
	\includegraphics[width=0.8\textwidth]{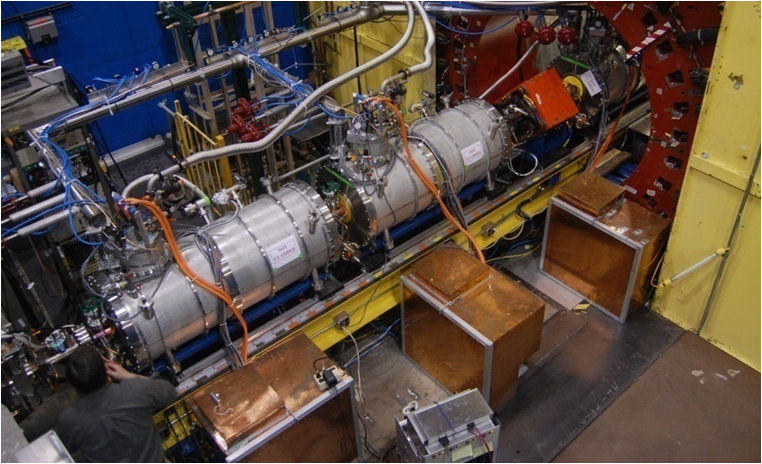}
	\caption[Superconducting wigglers in the L0 straight]{Superconducting wigglers in the L0 straight. \label{fig:l0_cleo}}
\end{figure}

In 2008, the L0 straight section of CESR was completely reconfigured for the CesrTA program.  The CLEO detector was removed, and six superconducting wigglers were installed (Fig.~\ref{fig:l0_cleo}).  The wigglers are 8-pole super-ferric magnets with main period of 40 cm and trimming end poles, and were typically operated with a peak transverse field of 1.9~T, closely matching the ILC DR wiggler requirements.  Three of these wigglers were instrumented with RFAs (Section~\ref{ssec:wig_inst}).

\subsection{Cloud Mitigation}

Evaluating the effectiveness of cloud mitigations in different magnetic field environments is a major component of the \cesrta~ program. Mitigations tested include:

\begin{itemize}
    \item TiN coating~\cite{NIMA551:187to199,NIMA578:470to479,NIMA556:399to409,NIMA564:44to50}, which reduces the secondary emission yield (SEY) of the coated chamber.
    \item Longitudinal grooves~\cite{NIMA571:588to598,Suetsugu2009449,JAP104:104904,PRSTAB7:034401}, which (in a dipole field) intercept electrons produced on the wall before they can drift to the center of the chamber.
    \item A clearing electrode~\cite{PAC07:Kryoer,NIMA598:372to378,PhysRevLett.110.124801}, which pulls electrons toward the chamber wall, away from the beam.
\end{itemize}

\section{\label{sec:instrumentation} Instrumentation}

RFAs in each field region had to be specially designed to fit inside the narrow magnet apertures.  A thorough account of the design and construction of these detectors can be found in~\cite{CLNS:12:2084}; here we provide an overview.  The key parameters of each RFA type are listed in Table~\ref{tab:dipole_rfa_styles}.

\begin{table}
   \centering
   \caption{\label{tab:dipole_rfa_styles}  RFA styles deployed in magnetic fields in CESR.  See~\cite{NIMA621:33to38} for more information on the chicane dipole RFAs.}
   \begin{tabular}{cccc}
       \hline \hline
            Type            &   No. of Grids   &   No. of Collectors    &   Grid Type       \\
       \hline
           CESR dipole      &   1       &   9       &   Etched   \\
           Chicane dipole   &   3       &   17      &   Mesh  \\
           Quadrupole       &   1       &   12      &   Mesh   \\
           Wiggler          &   1       &   12      &   Etched, Mesh \\
        \hline \hline
   \end{tabular}
\end{table}


\subsection{\label{ssec:grid_styles} General Design Considerations}

With the exception of the chicane dipole RFAs, which were designed at SLAC~\cite{NIMA621:33to38}, the detectors detailed below shared several design features:

\begin{itemize}
    \item A 3:1 depth to diameter ratio was chosen for the vacuum chamber holes to shield the detectors from direct beam signal.


    \item Because of the narrow apertures available inside the CESR magnets, only a single retarding grid was used, with the vacuum chamber serving as ground.

    \item The electron collector pads were laid out on copper-clad Kapton sheet using standard printed circuit board fabrication techniques.

\end{itemize}

Two different styles of retarding grid were used in the CESR RFAs (Table~\ref{tab:grid_styles}).  The first type was made of photo-chemically etched 0.15~mm-thick stainless steel (SST), with an optical transparency of approximately 38\%.  The grid holes have a bi-conical structure, with 0.18~mm diameter and 0.25~mm spacing.  To reduce secondary emission, the grids were coated with approximately 0.3~$\mu$m of gold.

The second style of grids was made with electroformed copper meshes, consisting of 15~$\mu$m wide and 13~$\mu$m thick wires with 0.34~mm spacing in both transverse directions.  They have an optical transparency of approximately 92\%.  The meshes were also coated with gold ($\sim$ 0.3~$\mu$m) via electroplating.  Both grid styles were used in the wiggler RFAs; Section~\ref{ssec:wig_inst} below discusses the merits of each style.

\begin{table}
   \centering
   \small
   \caption{\label{tab:grid_styles} Grid types used in CESR RFAs.  Note that ``transparency" refers to the optical transparency.}
   \begin{tabular}{cccc}
       \hline \hline
       Type &    Transparency   &   Material  &   Thickness \\
       \hline
           Etched   &   ~38\%   &   Gold coated SST   &   150~$\mu$m   \\
           Mesh   &   ~92\%   &   Gold coated Cu  &  13~$\mu$m   \\
        \hline \hline
   \end{tabular}
\end{table}

\subsection{\label{sssec:cesr_dip} CESR Dipole RFA}

\begin{figure}
	\centering
    \includegraphics[width=0.6\linewidth, angle=-90]{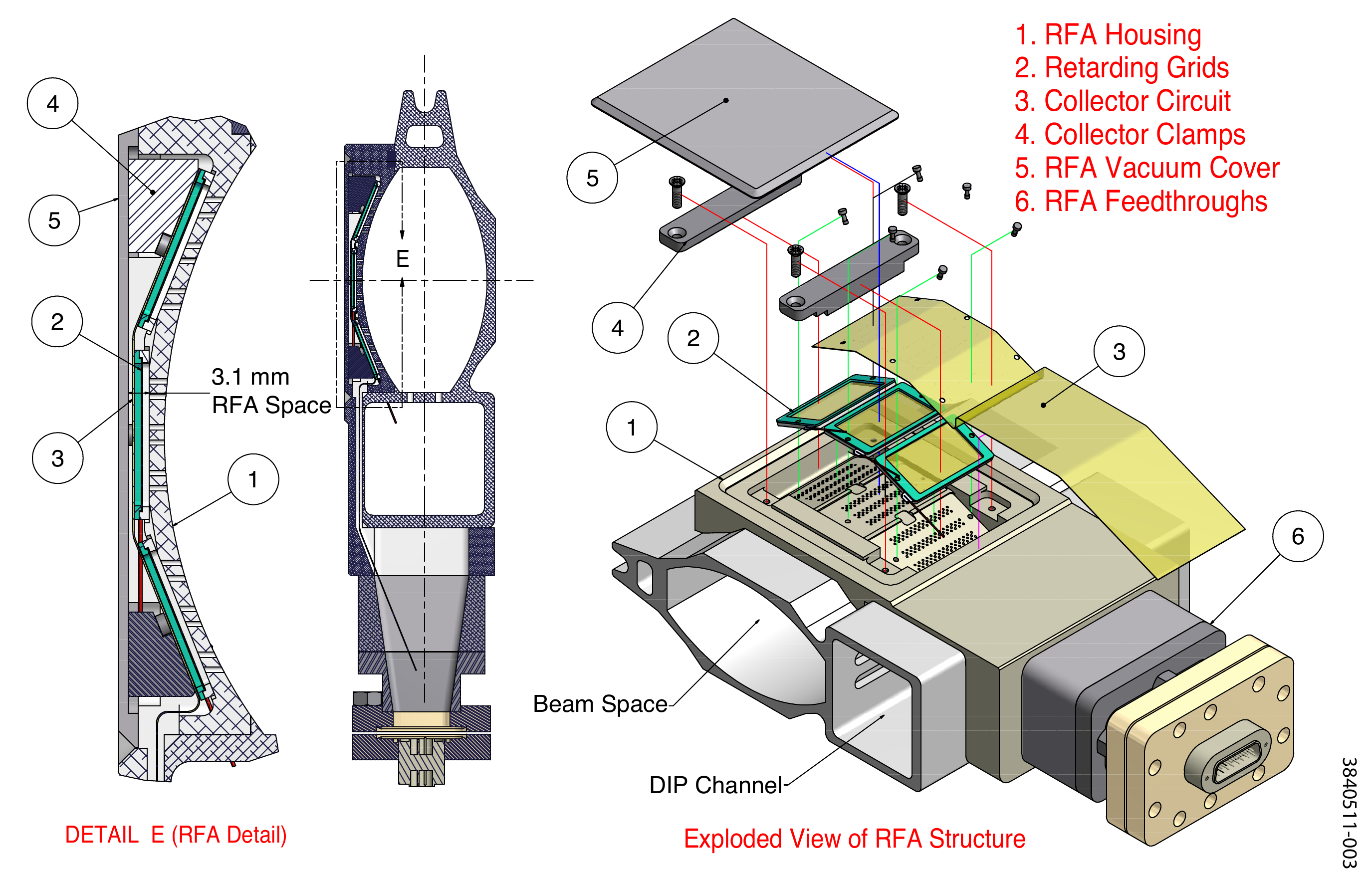}
	\caption{RFA design detail for a CESR dipole chamber. \label{fig:B12W_RFA}}	
\end{figure}

This RFA style was specifically designed for use inside the 12W CESR dipole, with only 3~mm available between the magnet and beam aperture.  As shown in Fig.~\ref{fig:B12W_RFA}, the RFA housing is machined from a separate block of 6061 aluminum welded to the cutout on top of the beam pipe.  The lower face of the RFA housing matches the curvature of the beam pipe aperture, while the upper face is divided into three flat sections, with each section tilted so that the wall thickness is similar.  There are three collectors in each section, for a total of nine.  Each section has a single retarding grid, made of etched stainless steel.  Arrays of 0.75~mm holes were drilled through the 2.5~mm thick beam pipe wall (maintaining the $\sim$3:1 ratio described above).  There are 44 holes per collector.  This style of RFA was also deployed in a field free region~\cite{NIMA760:86to97}.

\subsection{\label{sssec:chic_dip} Chicane Dipole RFAs}

\begin{figure}
	\centering
    \begin{tabular}{c}
	\includegraphics[width=0.9\textwidth]{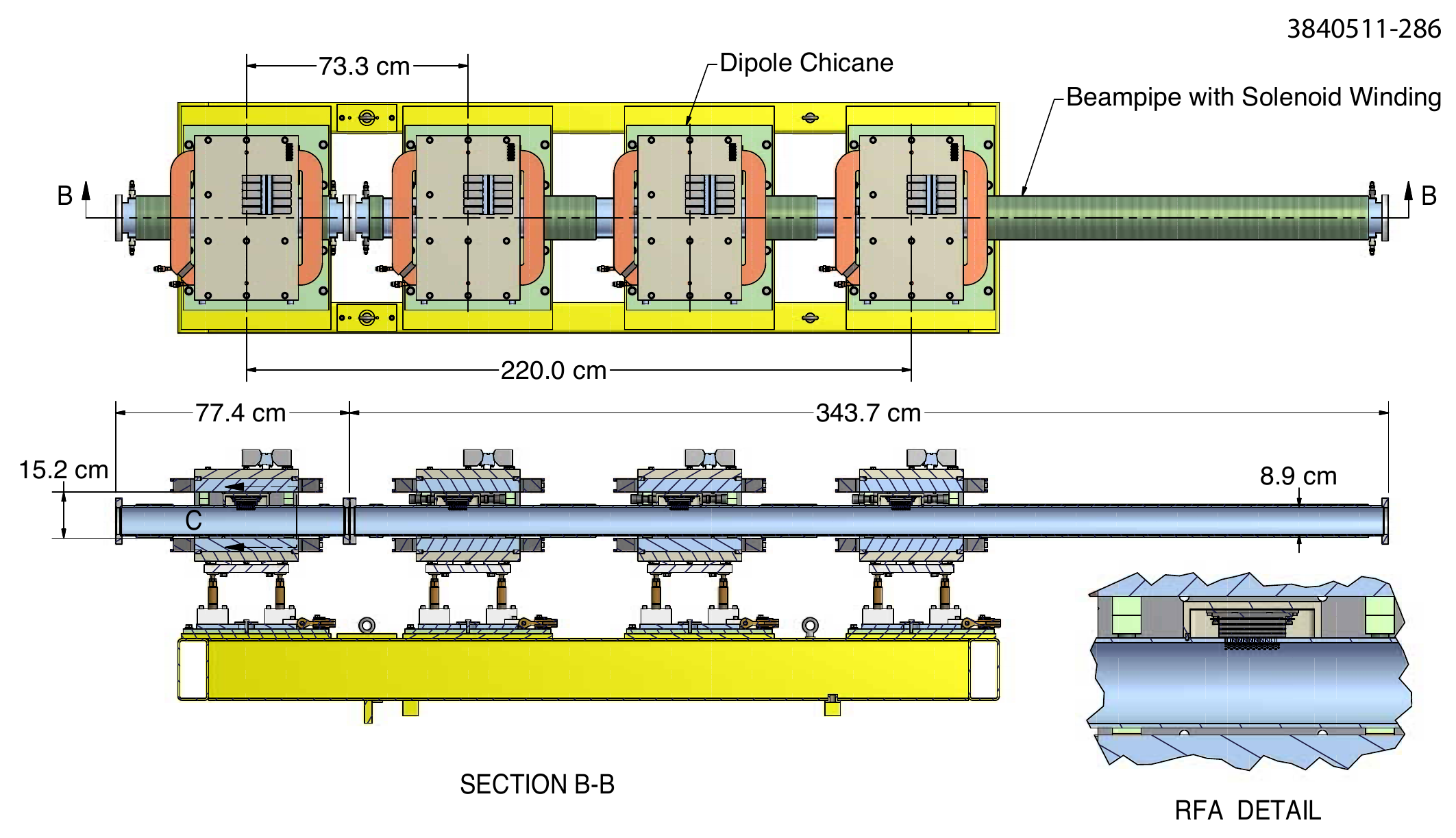}
    \end{tabular}
	\caption{PEP-II 4-dipole magnet chicane and RFA-equipped vacuum chambers. Top: top view.  Bottom: side view. \label{fig:cesr_conversion:SLAC_Chicane}}	
\end{figure}

The design of the 4 dipoles of the SLAC chicane is shown in Fig.~\ref{fig:cesr_conversion:SLAC_Chicane}. The chicane beam pipe is partitioned into 4 vacuum chambers for testing different electron cloud mitigation techniques.  One chamber is bare aluminum, two are TiN coated, and one is both grooved and TiN coated.  The grooves are triangular with a depth of 5.6~mm and an angle of 20$\degree$.  Each test chamber is equipped with an RFA and is surrounded by a chicane magnet.

\begin{figure}
	\centering
\begin{tabular}{cc}
\includegraphics[width=0.45\textwidth]{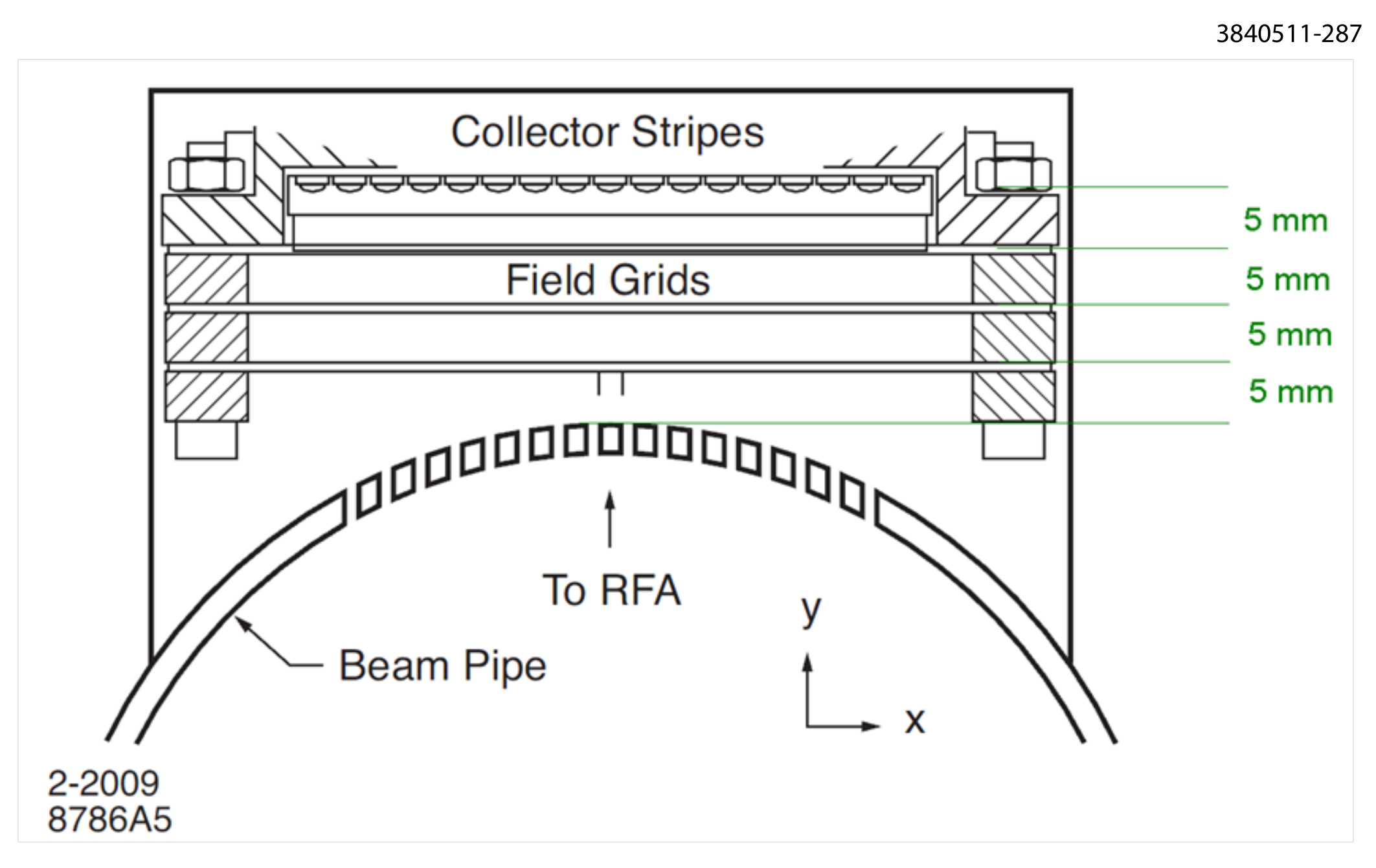} &
\includegraphics[width=0.45\textwidth]{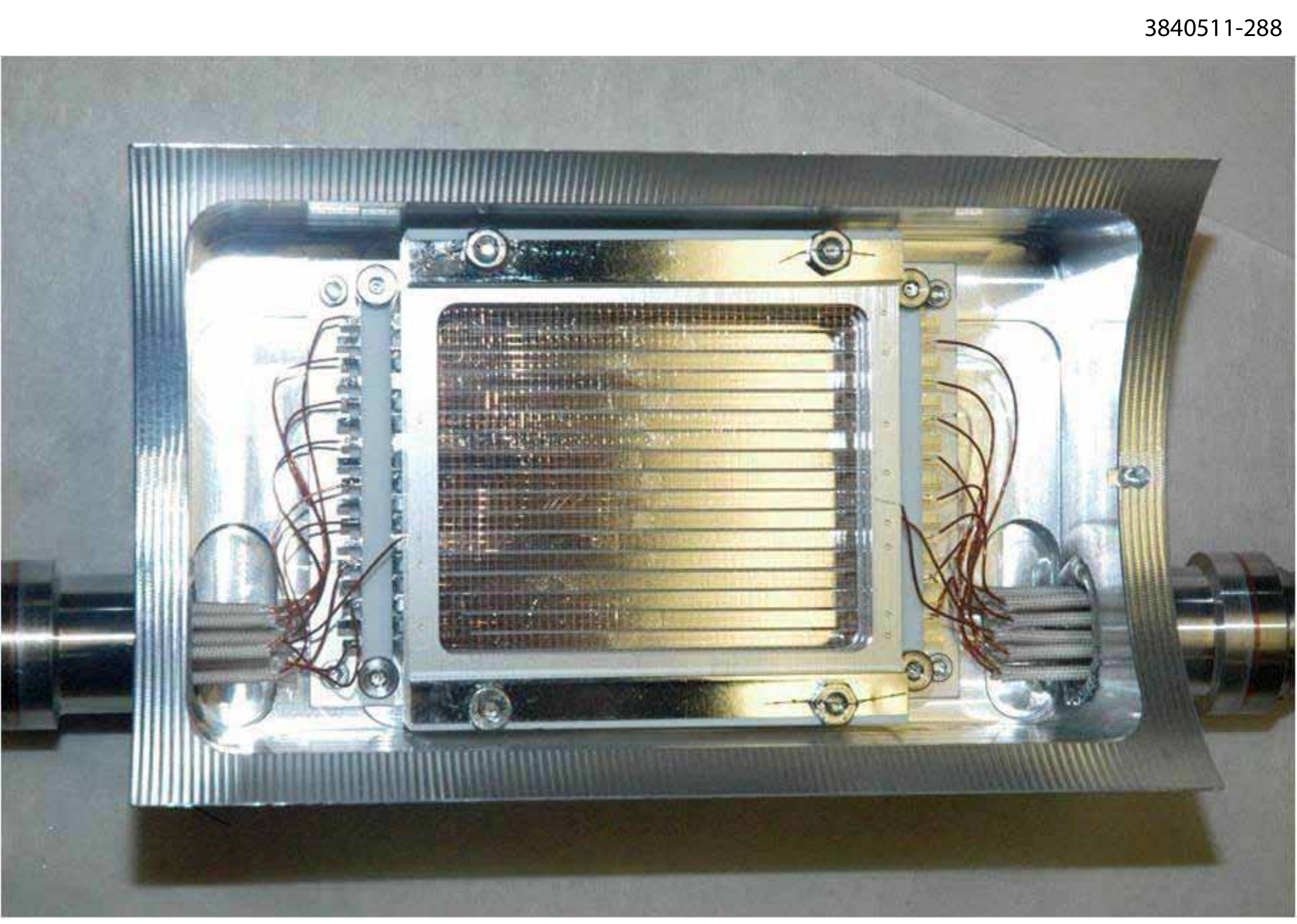}\\
\end{tabular}
	\caption{Four RFAs were welded onto the chicane beam pipes.  Left: Cross-sectional view.  Right: photograph of the assembled RFA in its aluminum housing.
	\label{fig:cesr_conversion:Chicane_RFA}}
\end{figure}

Figure~\ref{fig:cesr_conversion:Chicane_RFA} shows the structure of the chicane RFAs.  Because there was no aperture limitation for these magnets, the RFAs were designed with three (high efficiency) grids, with a generous 5~mm spacing between each grid.  The retarding voltage is applied to the middle grid.  The field of the chicane dipoles can be varied from 0 to 1.46~kG, though most of our measurements were done in a nominal dipole field of 810~gauss.


\subsection{\label{ssec:quad_inst} Quadrupole Instrumentation}

\begin{figure}
	\centering
    \includegraphics[width=0.6\linewidth, angle=-90]{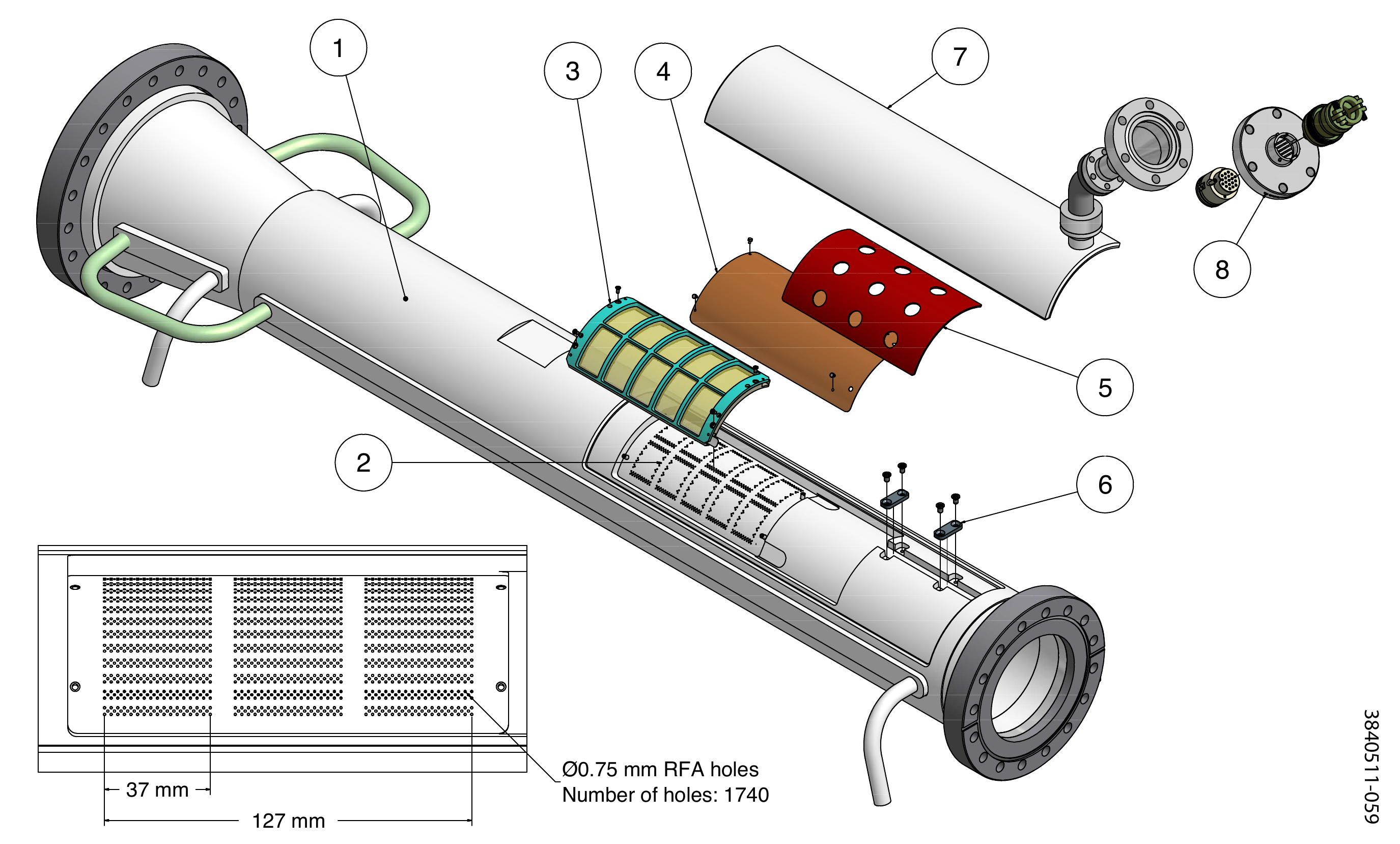}
	\caption[Exploded view of a RFA beam pipe in a CESR quadrupole magnet]{Exploded view of the structure of the RFA within a CESR quadrupole beam pipe. Major components include: (1) Aluminum beam pipe with cooling channels; (2) RFA housing and wiring channels; (3) Retarding grids, consisting of high-transparency gold-coated meshes; (4) RFA collector flexible circuit; (5) Stainless steel backing plate; (6) Wire clamps; (7) RFA vacuum cover with connection port; (8) Electric feedthrough. \label{fig:Quad_RFA_Structure}}	
\end{figure}
\begin{figure}
	\centering
	\includegraphics[width=0.75\textwidth, angle=-90]{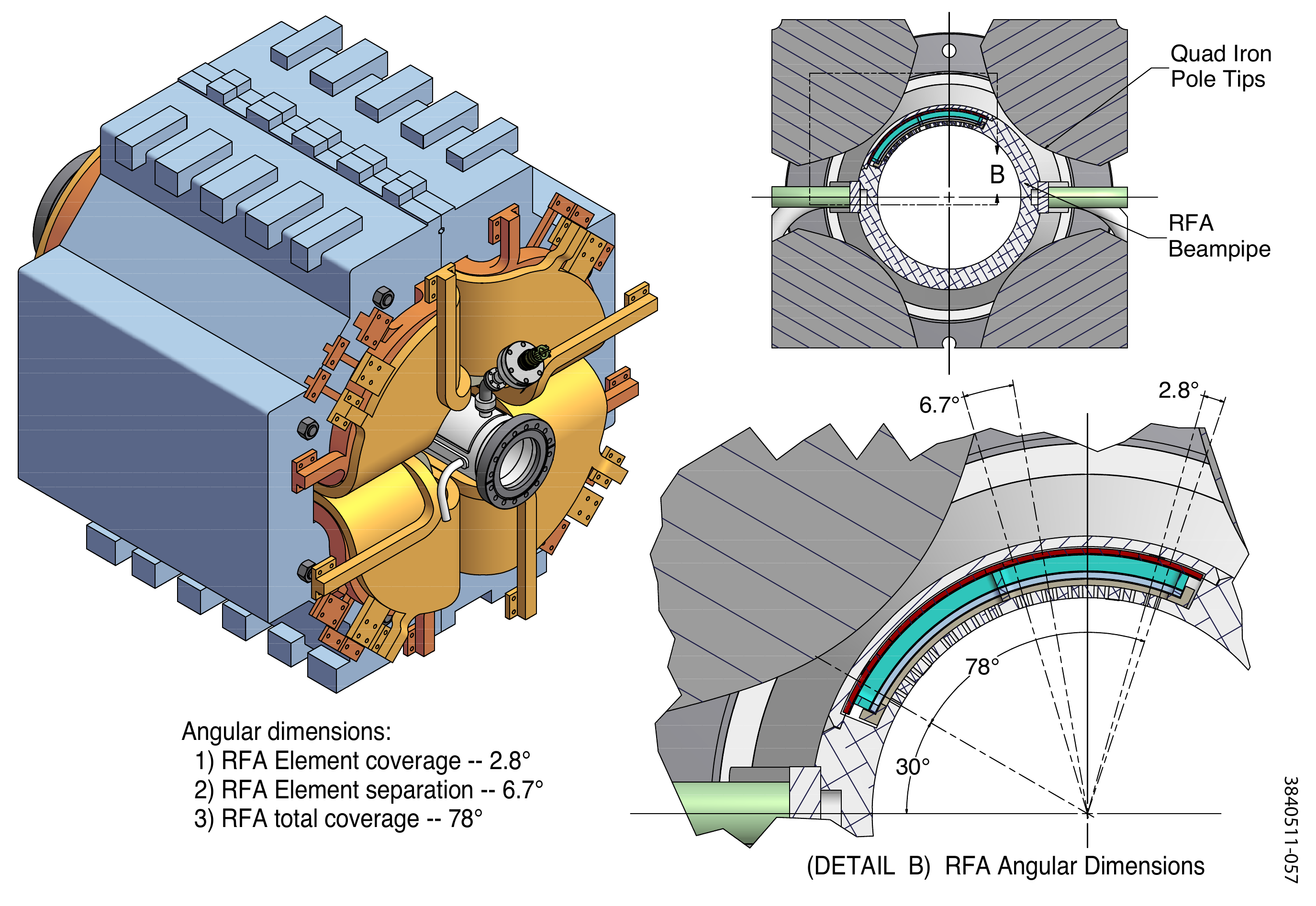}
	\caption{Left: Quadrupole with RFA beam pipe (left). Right: RFA angular coverage.\label{fig:RFA_in_quad}}
\end{figure}

The design of the quadrupole RFA beam pipe is illustrated in Figure~\ref{fig:Quad_RFA_Structure}.  This detector follows the curvature of the circular beam pipe, covering 78$^\circ$, including one of the quad pole tips.  There are 1740 beam pipe holes, which are grouped into 12~angular segments, matching the 12 RFA collector elements on the flexible circuit.  The beam-pipe-quadrupole assembly and angular coverage of the RFA are shown in Figure~\ref{fig:RFA_in_quad}.


\subsection{\label{ssec:wig_inst} Wiggler Instrumentation}

\begin{figure*}
	\centering
	\includegraphics[width=0.85\textwidth]{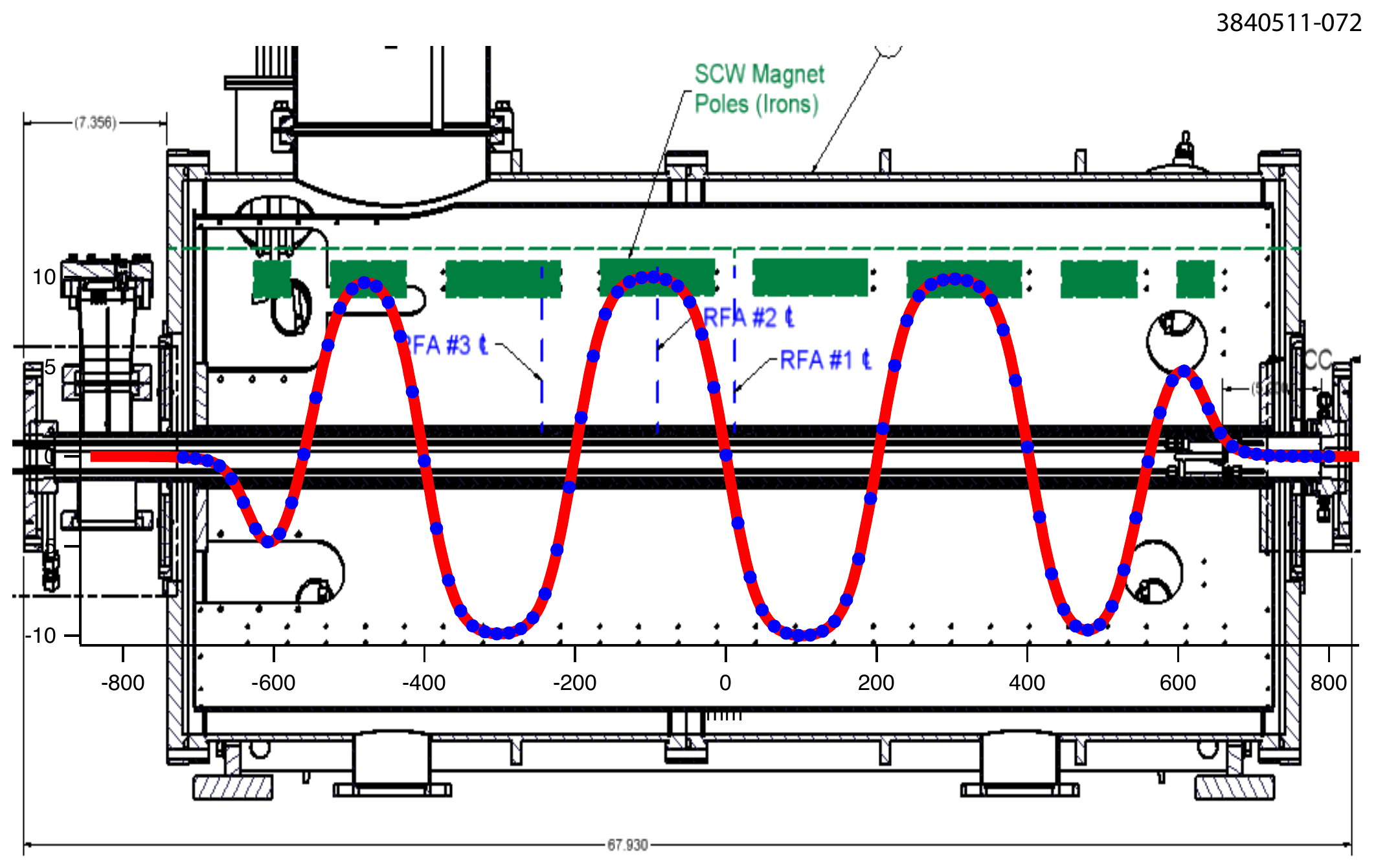}
	\caption[Locations of RFAs within the wiggler]{Three RFAs are built into each wiggler RFA beam pipe.  A plot of the magnetic field along the wiggler (red line with blue dots) is superimposed on the drawing of the wiggler.  The RFAs are located at three strategic locations. \label{SCW_RFA_locations}}
\end{figure*}

Three of the L0 wigglers (Section~\ref{sssec:l0}) were equipped with RFAs.  Each instrumented wiggler has three detectors: one in the center of a wiggler pole (where the field is mostly transverse), one in between poles (where the field is mostly longitudinal), and one in an intermediate region (where the field is manifestly three dimensional, see Fig.~\ref{SCW_RFA_locations}).  The RFAs each have one retarding grid and twelve collectors; there are 240 beam pipe holes per detector.

\begin{figure*}
	\centering
	\includegraphics[angle=-90, width=0.85\textwidth]{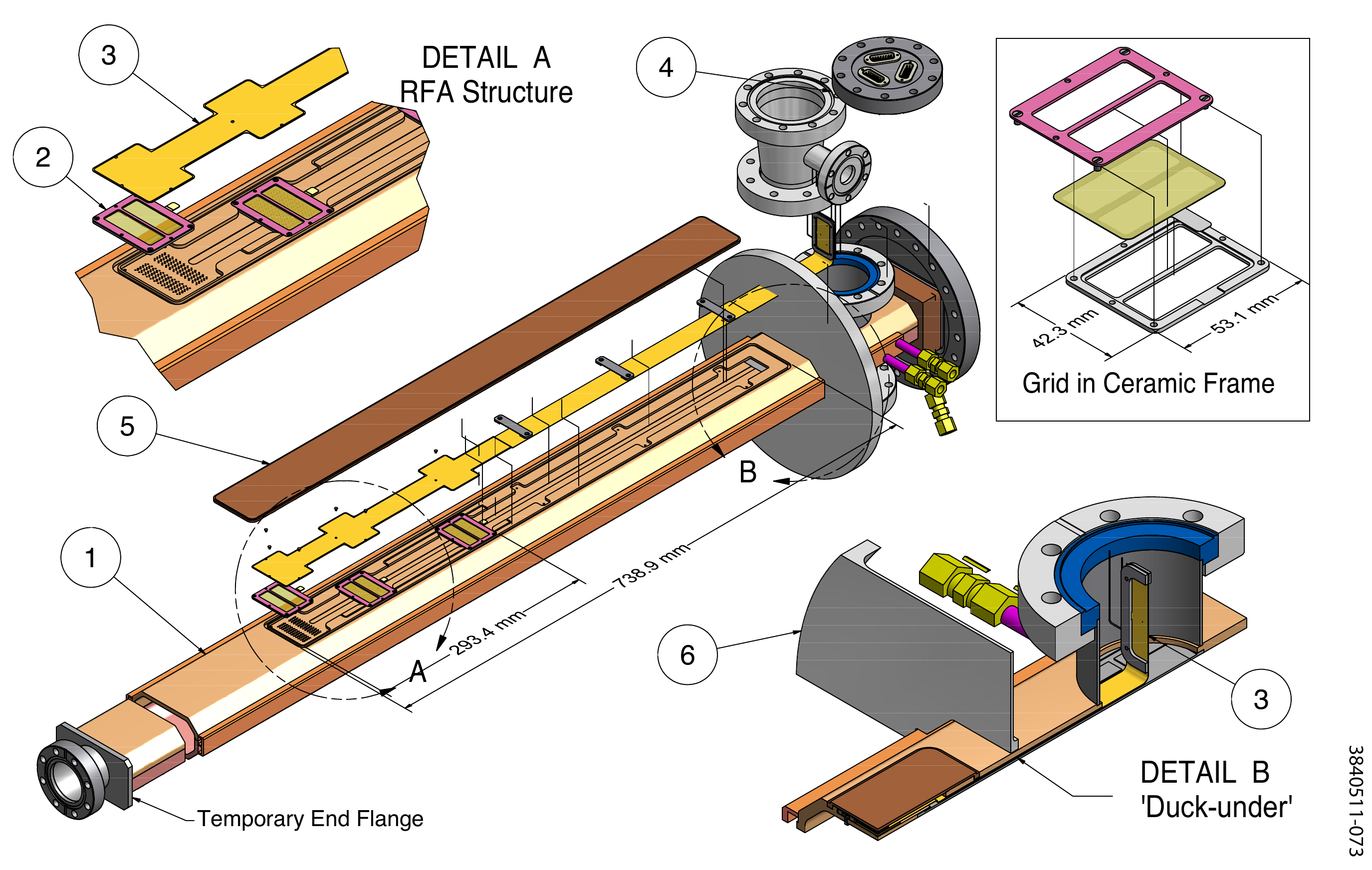}
	\caption[Exploded View of a SCW RFA beam pipe Assembly]{Exploded view of a wiggler RFA beam pipe assembly. The key components are: (1) beam pipe top half, housing the RFAs; (2) RFA grids (see upper right inset); (3) RFA collector; (4) RFA connection port; (5) RFA vacuum cover; (6) Flexible disk, which attaches the beam pipe to the wiggler vacuum vessel.  The ``duck-under" channel, through which the kapton flexible circuit is fed after all heavy welding is complete, is shown in Detail B.
\label{SCW_RFA_structure}}
\end{figure*}

The design of the wiggler RFAs was especially challenging.  An exploded view of the wiggler RFA beam pipe assembly is shown in Fig.~\ref{SCW_RFA_structure}.  Because the collector flexible circuit has a temperature rating of 220$\degree$ C, all the welding near the RFA had to be completed prior to the installation of the collector.  To allow for this requirement, a ``duck-under'' channel was created beneath the stainless steel flexible disk, which was later welded to the wiggler insulation vacuum vessel.  After all welding was complete (except for the final RFA vacuum cover), the collector flexible circuit was fed through the ``duck-under'' channel to the RFA portion of the beam pipe.  The construction of these chambers was done in collaboration with LBNL and KEK, and is described in more detail in~\cite{CLNS:12:2084} (Section 2.2.3.3).


The first generation wiggler RFAs were equipped with low-transparency stainless steel grids (see Table~\ref{tab:grid_styles}).  However, as described in Section~\ref{ssec:tramp}, secondary emission from these grids lead to a significant interaction between the electron cloud and the RFA, complicating the interpretation of the measurements.  Consequently, in the second generation of wiggler chambers, the grids were changed to high-transparency copper meshes.

The use of high efficiency grids effectively solved the grid emission problem.  However, in one of the RFAs (in the longitudinal field of the grooved chamber) the grid shorted to one of the collectors, preventing us from performing voltage scans with this detector.  Instead, we modified the electronics board so that the grid was powered by the collector power supply, to allow for passive data collection with the RFA.


Over the course of the \cesrta~ program, four different RFA-instrumented wiggler beam pipes have been constructed, each incorporating a different electron cloud mitigation mechanism.  The first two chambers, installed in fall 2008, were bare Cu and TiN coated Cu.  The TiN coating was applied (via DC sputtering) at SLAC.  In summer 2009, a chamber with triangular grooves on the bottom was constructed.  Finally, in 2010 we installed a chamber with a clearing electrode.  Additional information about the grooved and clearing electrode wiggler chambers can be found in~\ref{app:wig}.

These four chambers have been rotated through the three available locations in L0, to directly compare the effectiveness of each mitigation in identical photon environments and beam conditions.  The locations are nicknamed ``1W", ``2WA", and ``2WB" (ordered from upstream to downstream for the clockwise-travelling positron beam).  Table~\ref{wiggler_locs} gives the location of each instrumented wiggler chamber throughout the \cesrta~ program.  Note that in January 2011, the grooved chamber was coated with TiN, and installed at 2WA.


\begin{table}
    \centering
    \caption{Locations of instrumented wiggler chambers}
    \begin{tabular}{cccc}
      \hline \hline
      Dates & 1W & 2WA & 2WB  \\ \hline
      11/08 - 6/09 & - & Cu & TiN  \\
      7/09 - 3/10 & Cu & TiN & Grooves \\
      4/10 - 12/10 & Cu & TiN & Electrode \\
      1/11 - present & Cu & Grooves+TiN & Electrode \\
      \hline \hline
      \label{wiggler_locs}
    \end{tabular}
\end{table}

\section{\label{sec:measurements} Measurements}

Most of the data presented here are one of two types: ``voltage scans," in which the retarding voltage is varied (typically from +100 to $-250$~V or $-400$~V) while beam conditions are held constant, or ``current scans," in which the retarding grid is set to a positive voltage (typically +50~V), and data are passively collected while the beam current is increased.  The collector was set to +100~V for all of our measurements, to capture any secondary electrons produced on it.  The RFA signal is expressed in terms of current density in nA/mm$^2$, normalized to the geometric transparency of the RFA beam pipe and grids.  In principle, this gives the time averaged electron current density incident on the beam pipe wall.


\subsection{Dipole Measurements}

\begin{figure}
\centering
\begin{tabular}{c}
\includegraphics[width=.6\textwidth]{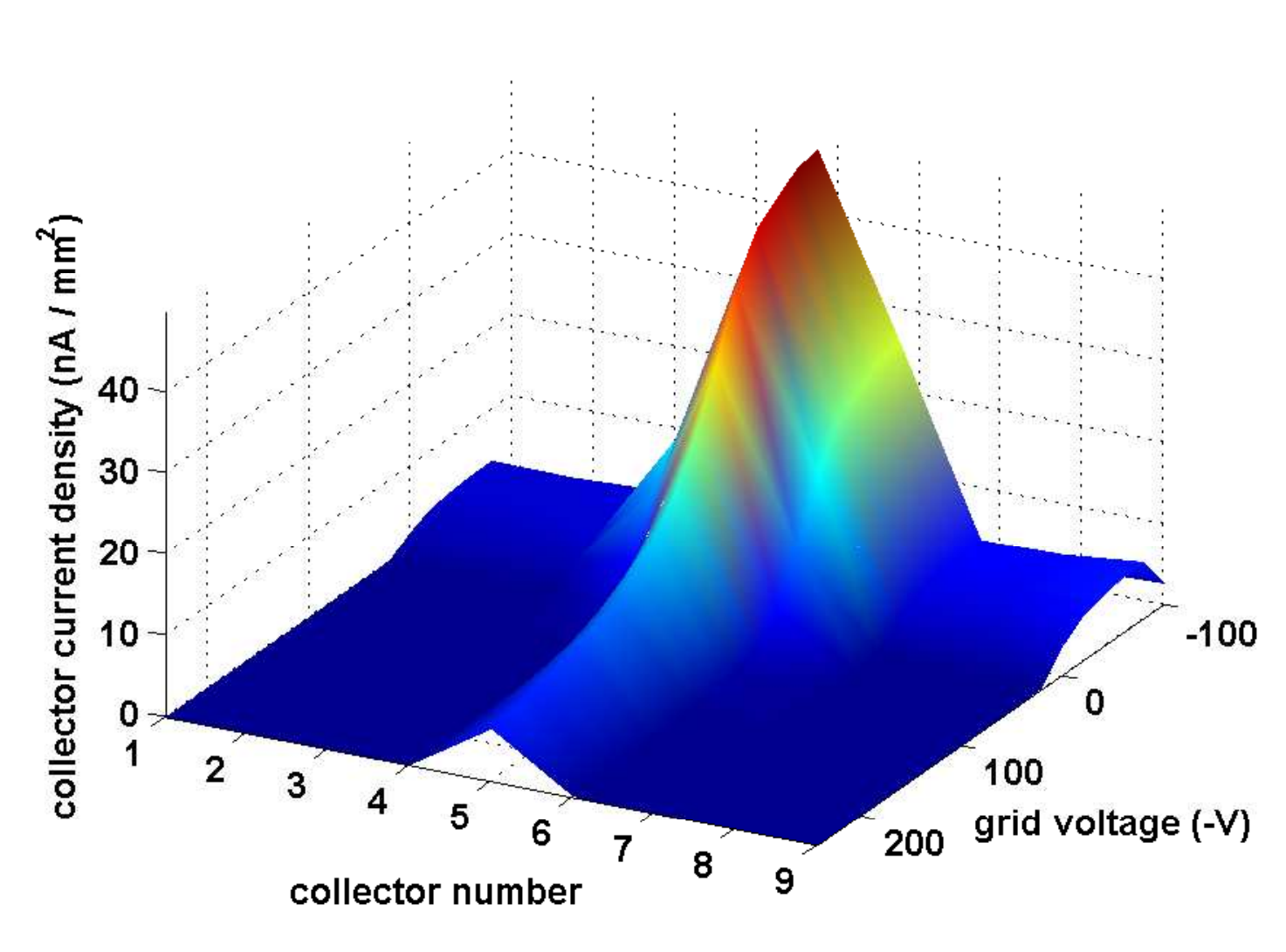} \\ 
\includegraphics[width=.6\textwidth]{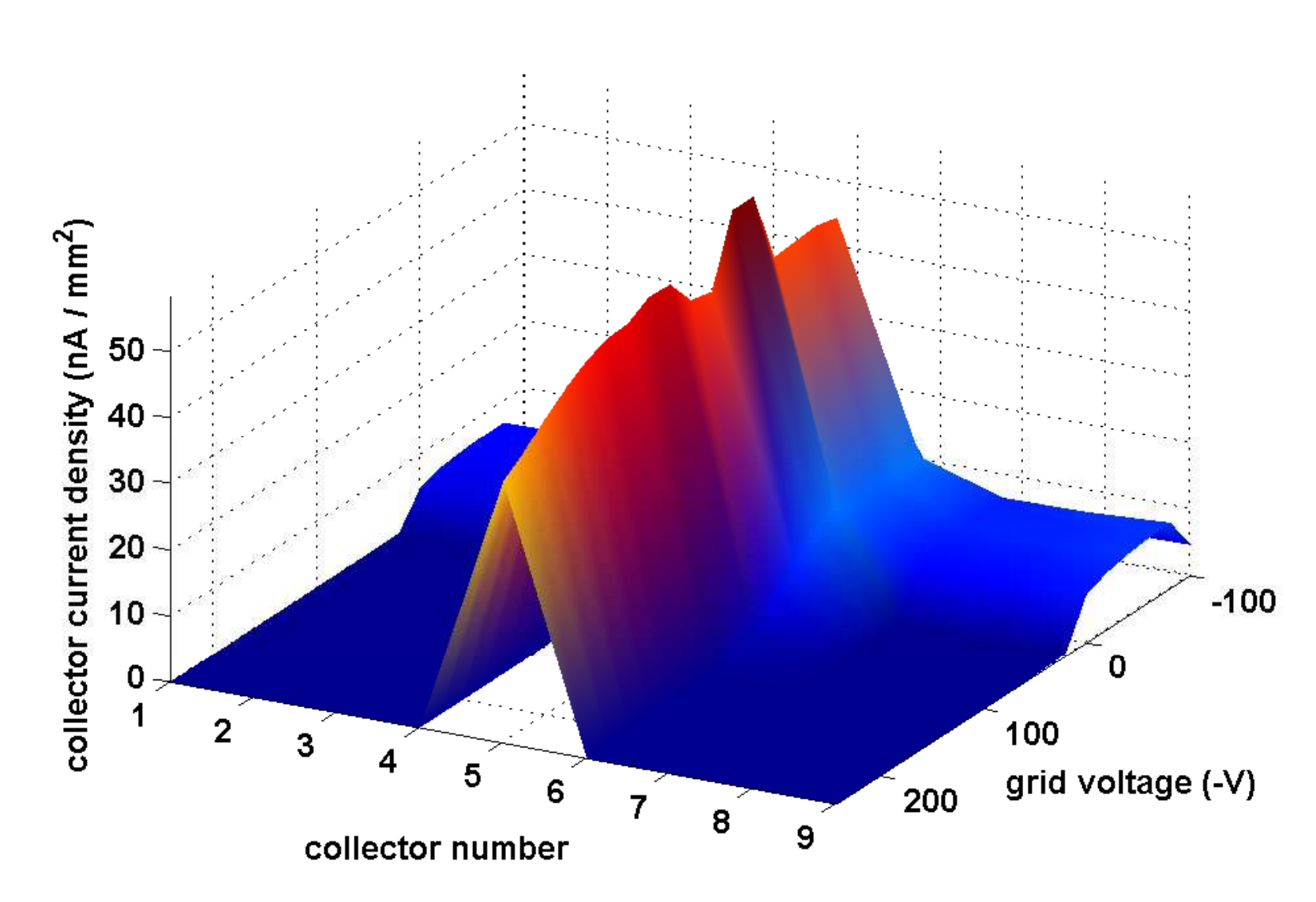} \\
\end{tabular}
\caption[CESR dipole RFA voltage scans]{\label{fig:cesr_dipole_meas} CESR dipole RFA voltage scans: 1x45x1.25~mA e+, 14ns.  Top: 2.1~GeV (790 gauss); Bottom: 5.3~GeV (2010 gauss). The central collector is no. 5, and the chamber is made of uncoated aluminum.}
\end{figure}

Fig.~\ref{fig:cesr_dipole_meas} shows retarding voltage scans done with the CESR dipole RFA under standard \cesrta~conditions, at two different beam energies.  The beam conditions are given as ``1x45x1.25~mA e$^+$, 14~ns."  This notation, which will be used throughout this section, indicates one train of 45 bunches, with 1.25~mA/bunch (1~mA = $1.6\times10^{10}$ particles), with positrons and 14~ns spacing.  At both energies, one can see a strong multipacting peak in the central collector, aligned with the horizontal position of the beam.  This is where cloud electrons receive the strongest beam kicks, and generate the most secondaries.  The measured energy distribution is qualitatively very different for the two beam energies, an effect which was not observed in other detectors.  This discrepancy is due to the fact that the RFA efficiency
depends on the magnetic field strength (which in turn depends on the beam energy in the CESR dipoles), as will be discussed in Section~\ref{ssec:dip_strength}.

\begin{figure}
\centering
\begin{tabular}{c}
\includegraphics[width=.6\textwidth]{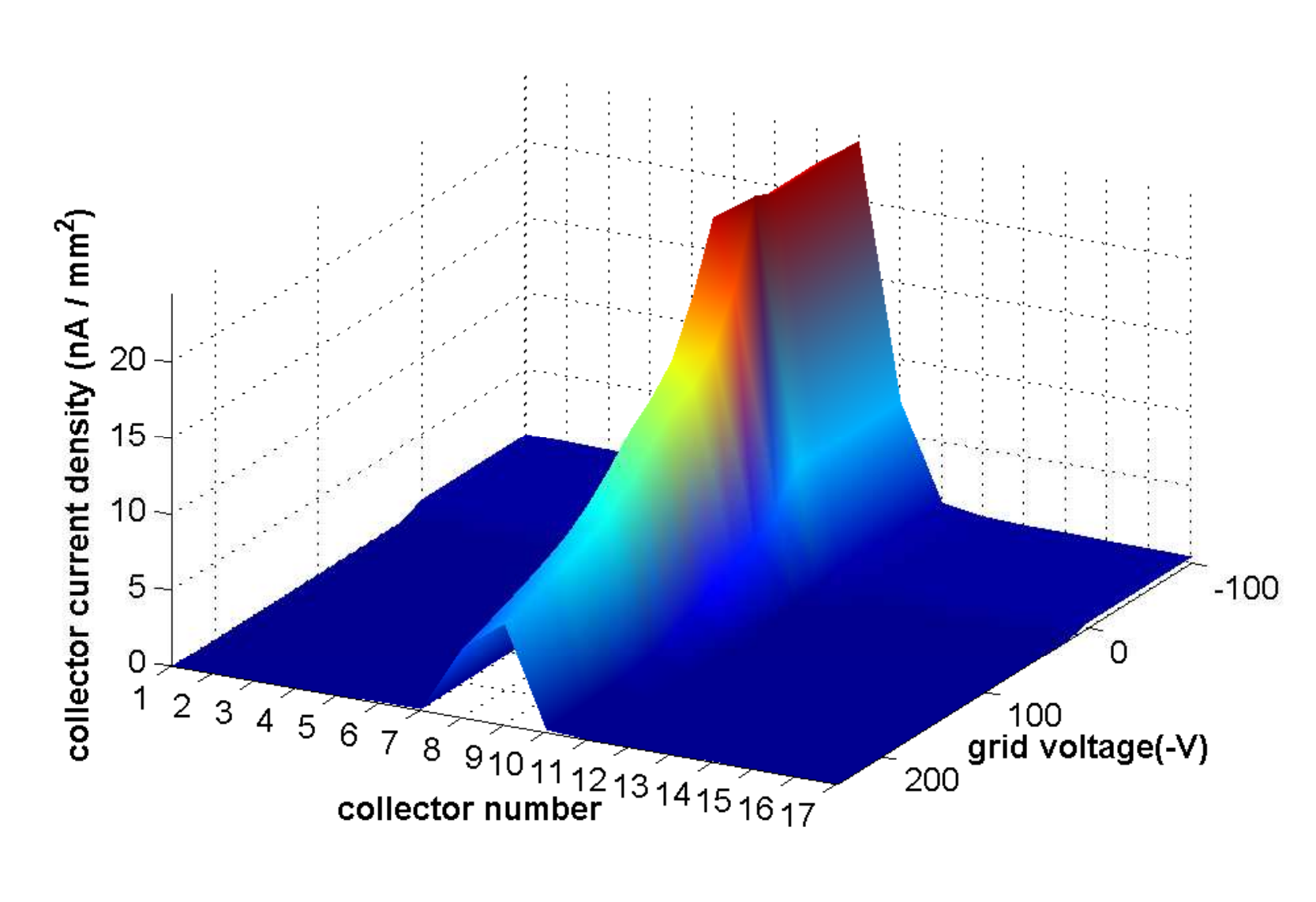} \\    
\includegraphics[width=.6\textwidth]{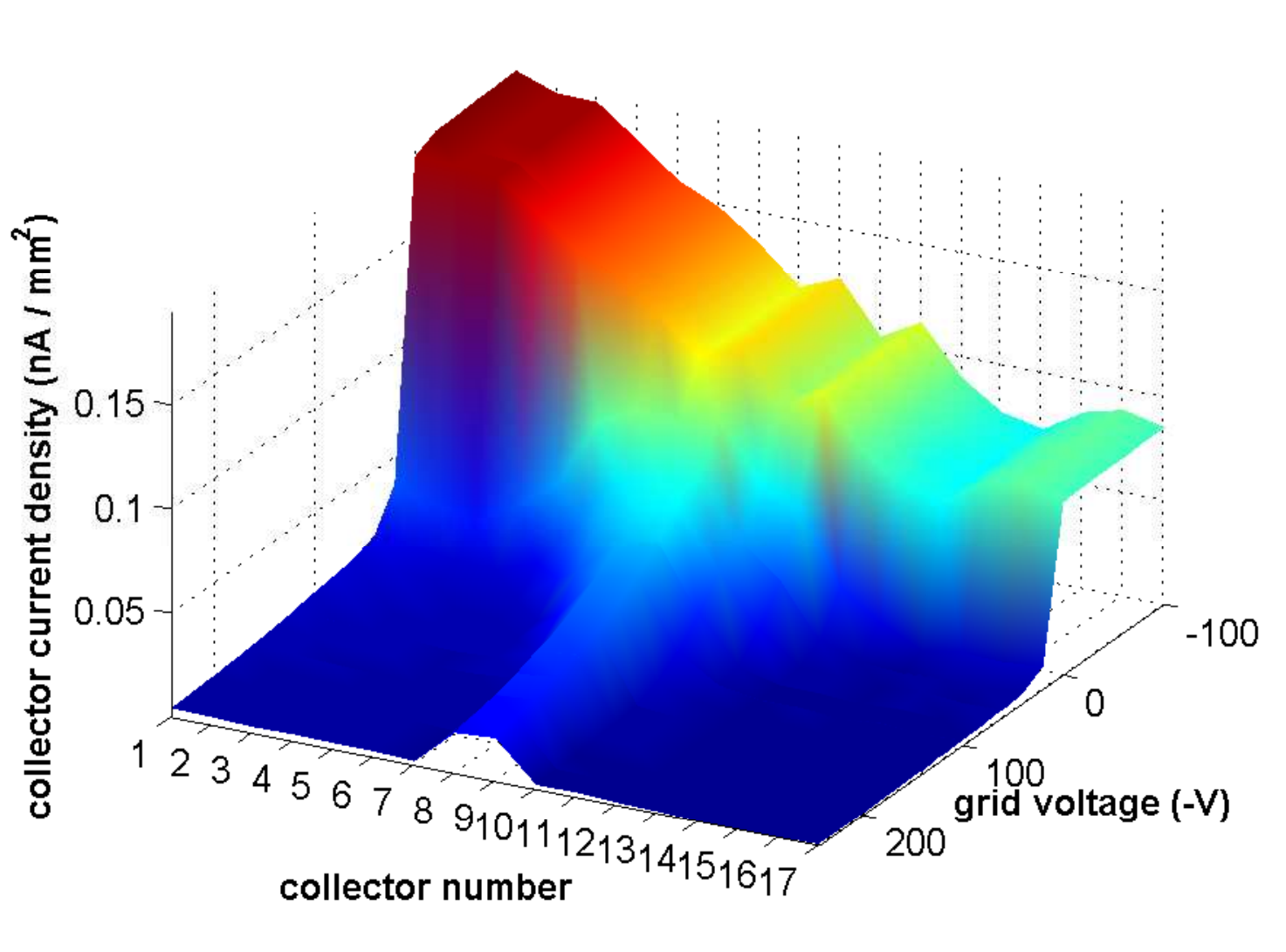} \\ 
\end{tabular}
\caption[CESR dipole RFA voltage scans]{\label{fig:chic_dipole_meas} Chicane dipole RFA voltage scans: 1x45x1.25~mA e+, 14~ns, 5.3~GeV, 810~gauss.  Top: Aluminum chamber; Bottom: TiN-coated chamber. The central collector is no. 9.}
\end{figure}

Voltage scans done under similar conditions in the chicane RFAs are shown in Fig.~\ref{fig:chic_dipole_meas}.  In the aluminum chamber, we again observe a strong multipacting peak.  However, in the TiN coated chamber, where the SEY is much lower, the peak is greatly suppressed, and the overall signal is much lower.  The variation in signal across the collectors in this measurement is most likely due to an azimuthal variation in the photon flux.

\subsubsection{\label{ssec:bifurcate} Bifurcation of Central Peak}

\begin{figure}
   \centering
   \includegraphics*[width=0.6\textwidth]{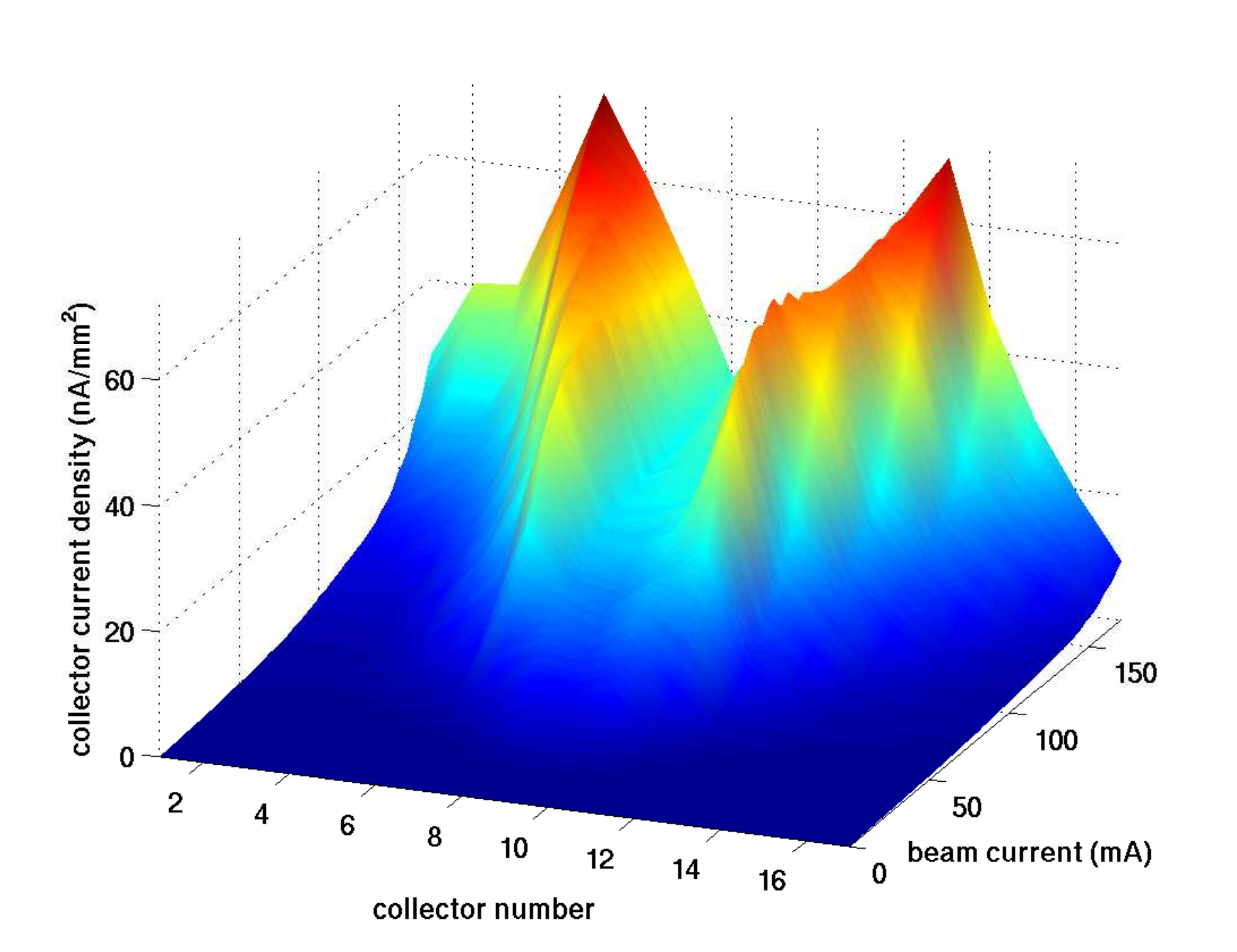}   
   \caption[Bifurcation of peak cloud density in a Al dipole]{\label{fig:bifurcate} Bifurcation of peak cloud density in a Al chicane dipole: 1x20 e+, 5.3~GeV, 14~ns.}
\end{figure}

\begin{figure}
	\centering
    \includegraphics[width=0.6\linewidth]{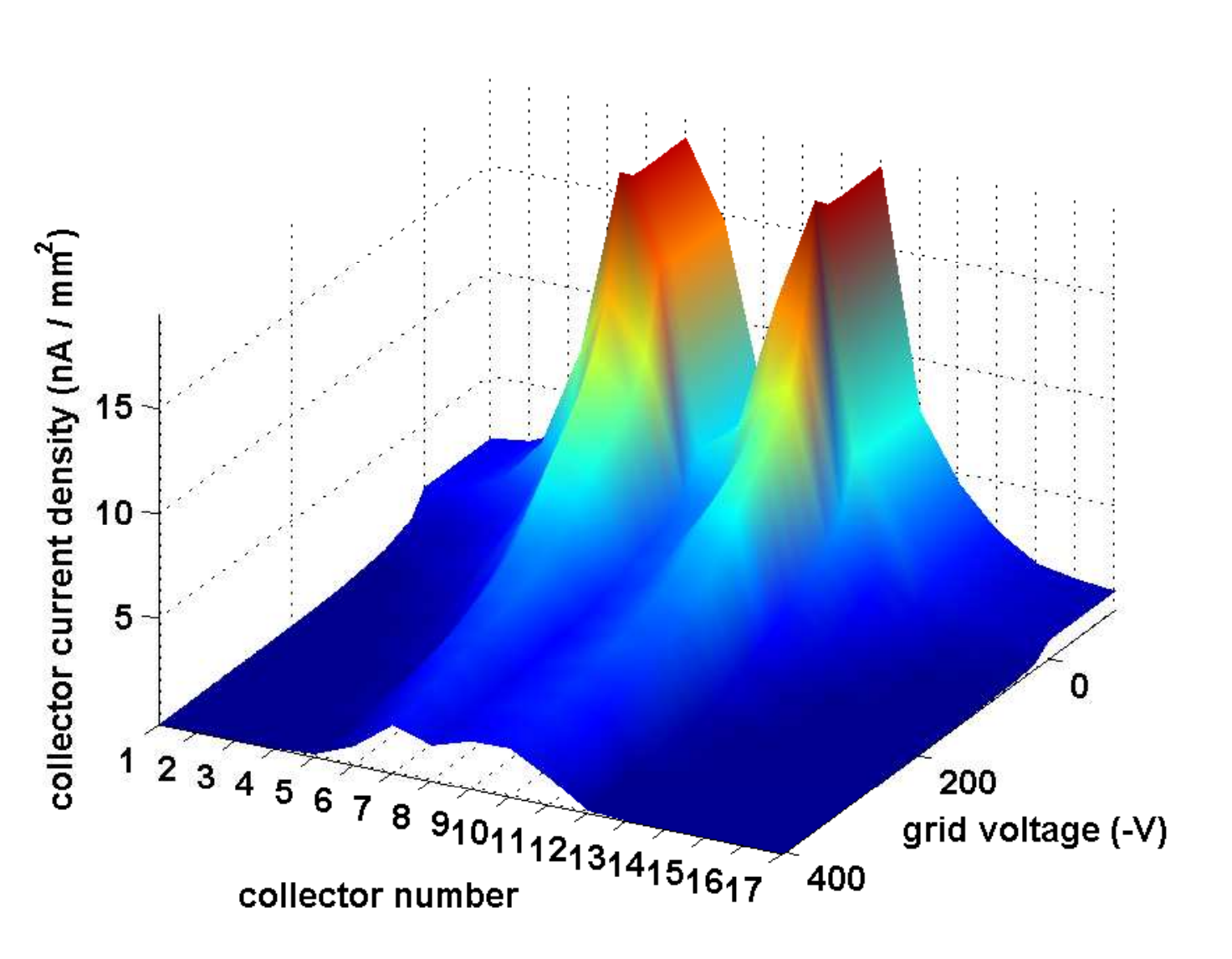}
	\caption[Chicane RFA voltage scan at high beam current]{Chicane RFA voltage scan at high beam current: 1x20x5~mA e+, 5.3~GeV, 14~ns.  Bifurcation is clearly visible. \label{fig:vscan_bifur}}	
\end{figure}

For high bunch currents, we have observed a bifurcation of the central multipacting peak into two peaks with a dip in the middle.  This is demonstrated in Fig \ref{fig:bifurcate}, which shows the signal in the aluminum chamber chicane RFA vs beam current.  Bifurcation occurs when the average energy of electrons in the center of the beam pipe is higher than the peak energy of the SEY curve, so that the location of the effective maximum yield is actually off center~\cite{NIMA621:33to38}.  The higher the bunch current, the further off center these peaks will be.  Voltage scans done with high beam current (see Fig.~\ref{fig:vscan_bifur}), also clearly show this effect.

\subsubsection{\label{ssec:dip_strength} Dependence of RFA Signal on Dipole Field Strength}

\begin{figure}
    \centering
    \begin{tabular}{c}
    \includegraphics[width=.6\textwidth]{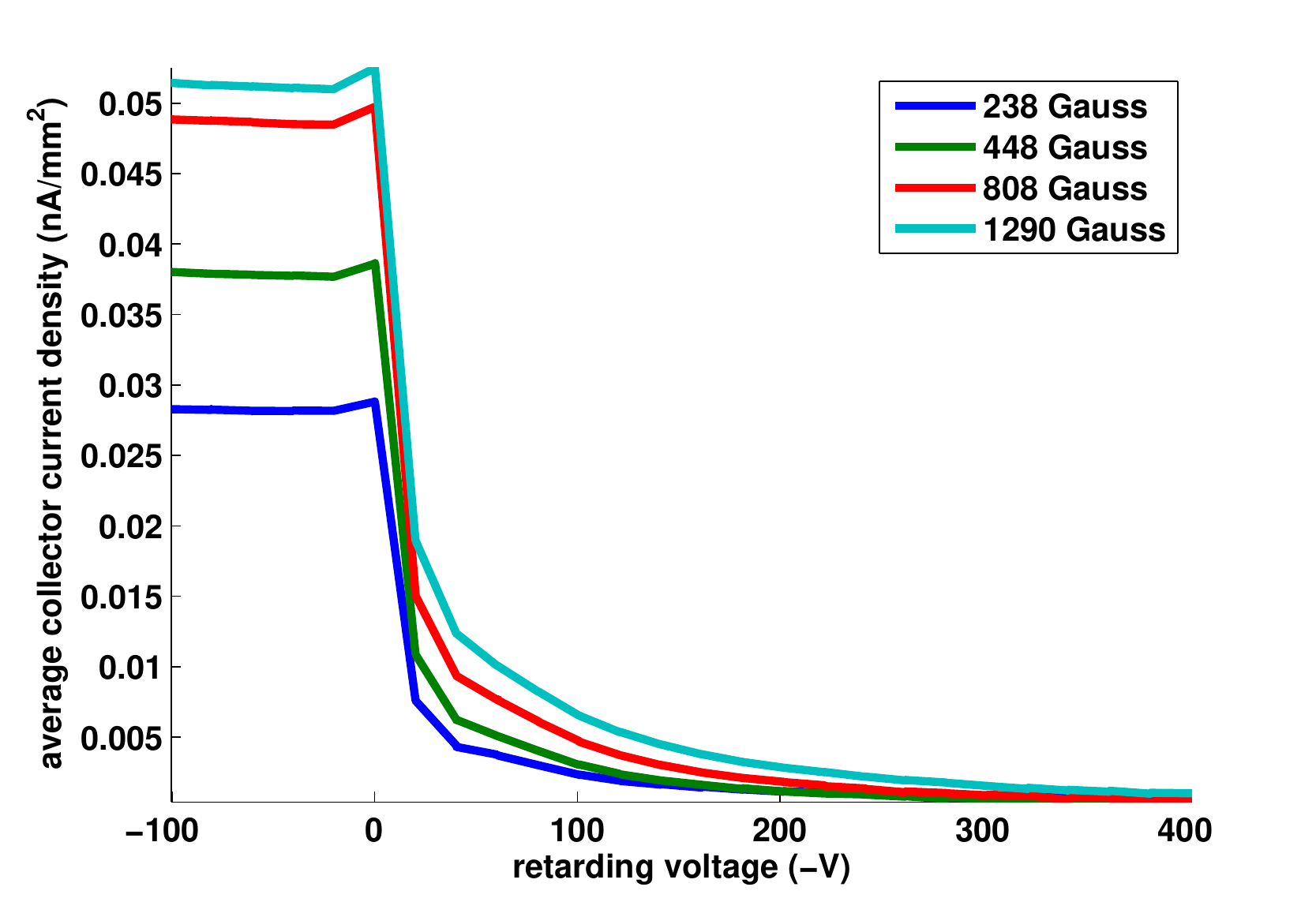} \\
    \includegraphics[width=.6\textwidth]{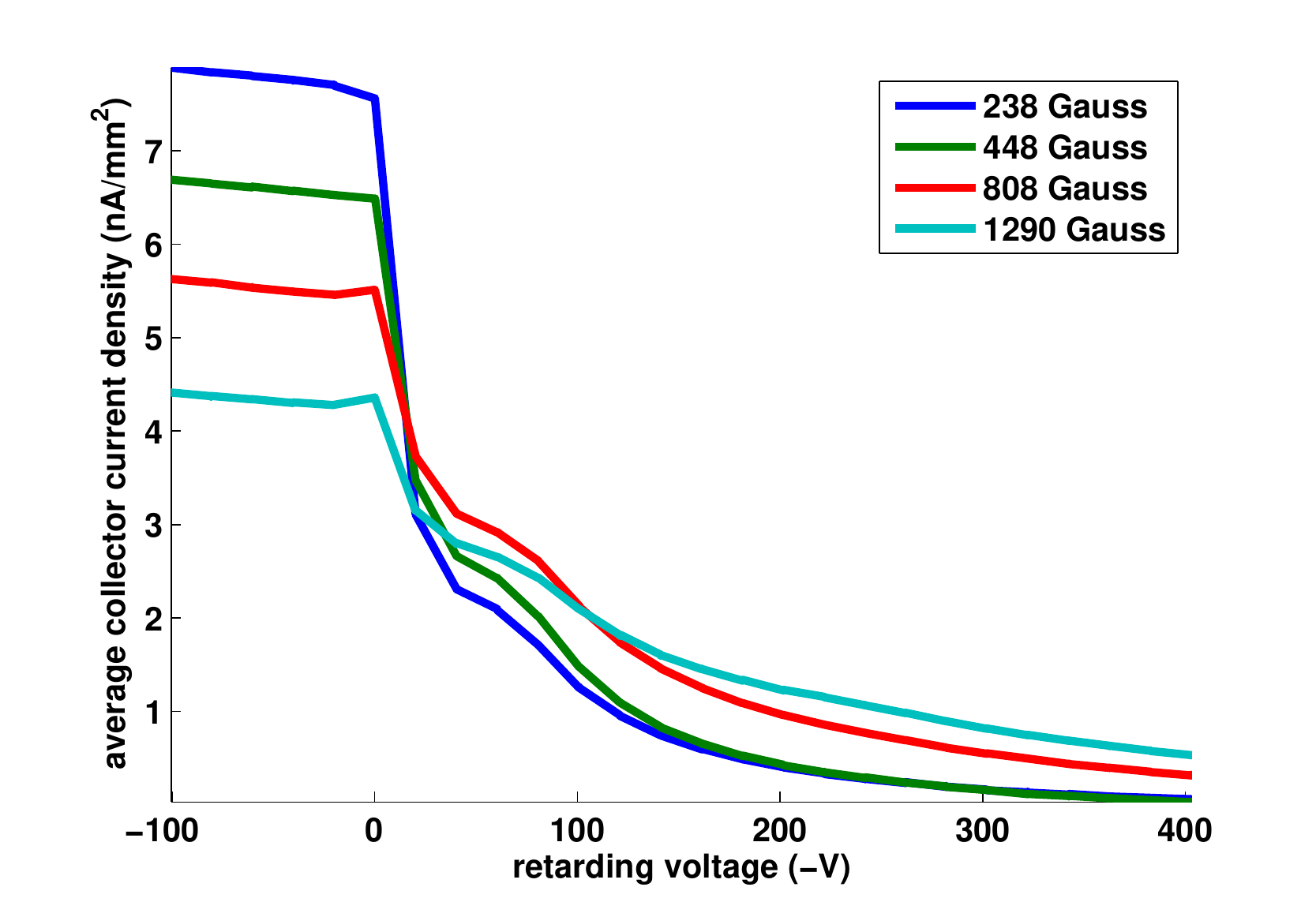} \\
    \end{tabular}
    \caption{\label{fig:chic_field_compare} RFA voltage scans for different chicane field strengths.  Top: TiN coated chamber.  Bottom: bare Al chamber.  Beam conditions are 1x20x2.8~mA e+, 14~ns, 5.3~GeV.}
\end{figure}

The strength of the magnetic field can have a significant effect on dipole RFA measurements.  Fig.~\ref{fig:chic_field_compare} shows a series of voltage scans performed with the aluminum chicane RFA, under the same beam conditions, with different dipole field values.  In the TiN coated chamber, the signal increases with field strength.  This is because as the field is increased, the cyclotron radius of the electrons becomes smaller, so they can more easily pass through the beam pipe holes.  This results in a higher effective RFA efficiency~\cite{IPAC12:WEPPR088}.  The same effect is observed in the aluminum chamber at high retarding voltage.  However, the total signal in the aluminum chamber actually decreases with field.  At very high fields, the motion of the electrons becomes almost completely one dimensional, and the RFA begins to deplete the cloud directly under the beam pipe holes.  This effectively makes the detector less sensitive to multipacting, since electrons will not be able to collide with the vacuum chamber multiple times before being absorbed.

Notably, simulations indicate that the cloud development is almost completely insensitive to field strength for the range of fields shown in Fig.~\ref{fig:chic_field_compare}.  This means that the difference in RFA signal is purely a detector effect.  More detailed simulations are required to confirm that the efficiency and depletion effects described above quantitatively explain what we measure.

\subsection{Quadrupole Measurements}

\begin{figure}
   \centering
   \includegraphics*[width=0.6\textwidth]{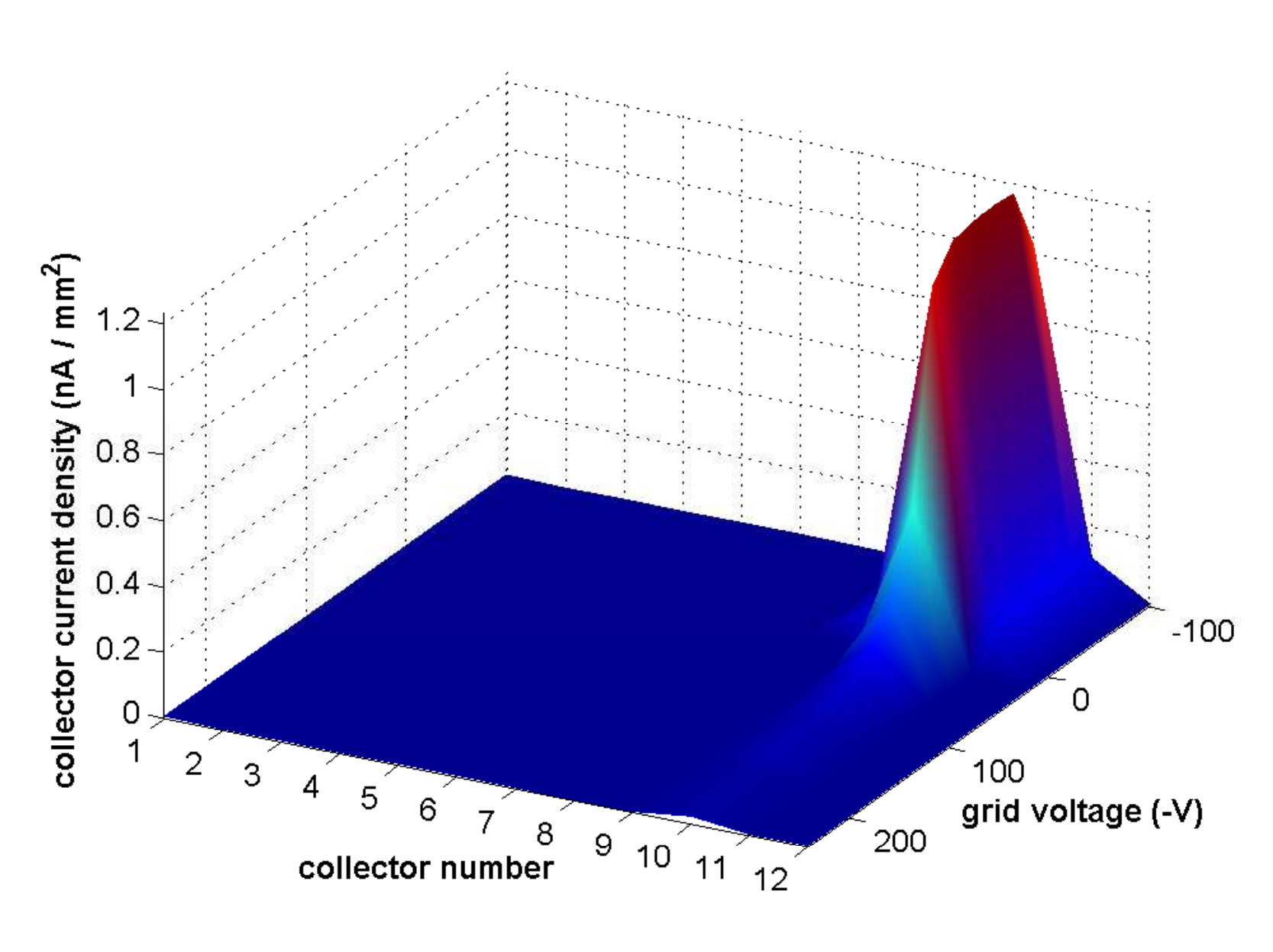}       
   \caption[Quadrupole RFA voltage scan]{\label{fig:quad_example} Quadrupole RFA voltage scan: 1x45x1.25~mA e+, 5.3~GeV, 14~ns.}
\end{figure}

A typical quadrupole RFA measurement is shown in Fig.~\ref{fig:quad_example}.  We observe high current on the collector lined up with the quad pole tip (no. 10), and little current on the other collectors.  This suggests that the majority of the cloud is streaming between the pole tips, guided by the quadrupole field lines.


One particular area of concern for quadrupole magnets is the potential for long term ($> 1$~$\mu$s) cloud trapping.  Since the RFA provides a time integrated measurement, it is not ideally suited for addressing this question, though data taken as a function of bunch spacing is consistent with a long time scale for cloud development~\cite{IPAC11:MOPS083}.  More direct evidence of quad trapping has come from measurements done with a time resolved electron cloud detector~\cite{NAPAC13:FROAA5}.

\subsection{\label{ssec:wig_meas} Wiggler Measurements}

\begin{figure}
   \centering
   \includegraphics*[width=0.6\textwidth]{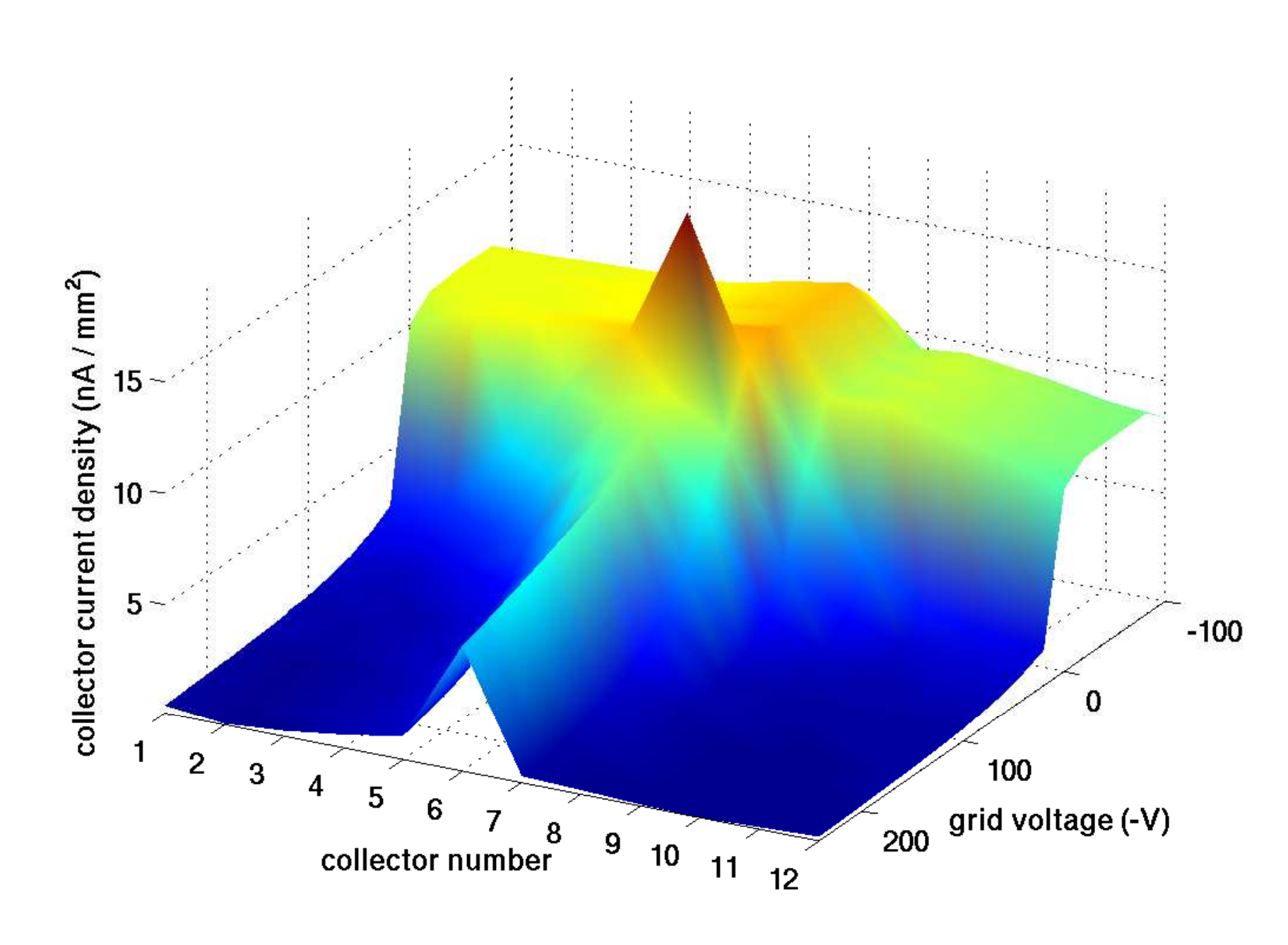}       
   \caption[Cu Wiggler RFA measurement]{\label{fig:cu_wig} Cu wiggler pole center RFA measurement: 1x45x1.25~mA e+, 2.1~GeV, 14~ns.}
\end{figure}

Fig.~\ref{fig:cu_wig} shows a typical voltage scan done in the center pole RFA (see Fig.~\ref{SCW_RFA_locations}) of a Cu wiggler chamber, for a 45 bunch train of positrons at 1.25~mA/bunch, 14~ns spacing, and 2.1 GeV.  The signal is fairly constant across all the collectors at low retarding voltage, but does become peaked at the center at high energy.  There is also an anomalous spike in current at low (but nonzero) retarding voltage.  Since the RFA should simply be collecting all electrons with an energy more than the magnitude of the retarding voltage, the signal should be a monotonically decreasing function of the voltage.  So the RFA is not behaving simply as a passive monitor.  This effect is discussed in detail in Section~\ref{ssec:tramp})

\begin{figure}
   \centering
   \includegraphics*[width=0.6\textwidth]{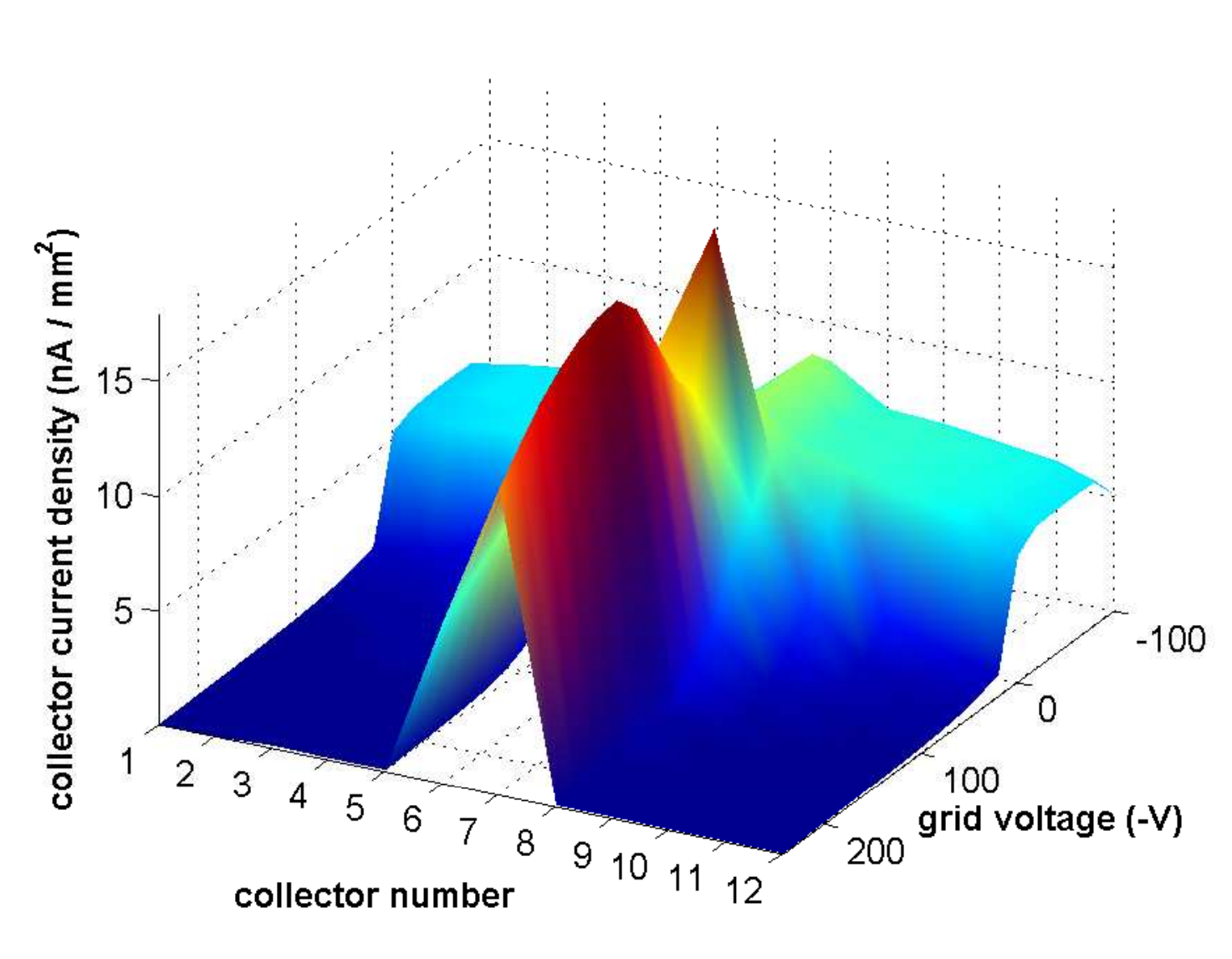} 
   \caption[Cu Wiggler RFA measurement]{\label{fig:cu_wig_int} Cu wiggler intermediate field RFA measurement: 1x45x1.25~mA e+, 2.1~GeV, 14~ns.}
\end{figure}

A voltage scan done in the intermediate field RFA under the same beam conditions is shown in Fig.~\ref{fig:cu_wig_int}.  Typical of measurements with this detector, we observe complex (and unphysical) behavior as a function of retarding voltage.  These measurements are not currently well understood.  Modeling of cloud development at this location (requiring a fully three dimensional simulation)  is a potential future area of research.

\begin{figure}
   \centering
   \includegraphics*[width=0.6\textwidth]{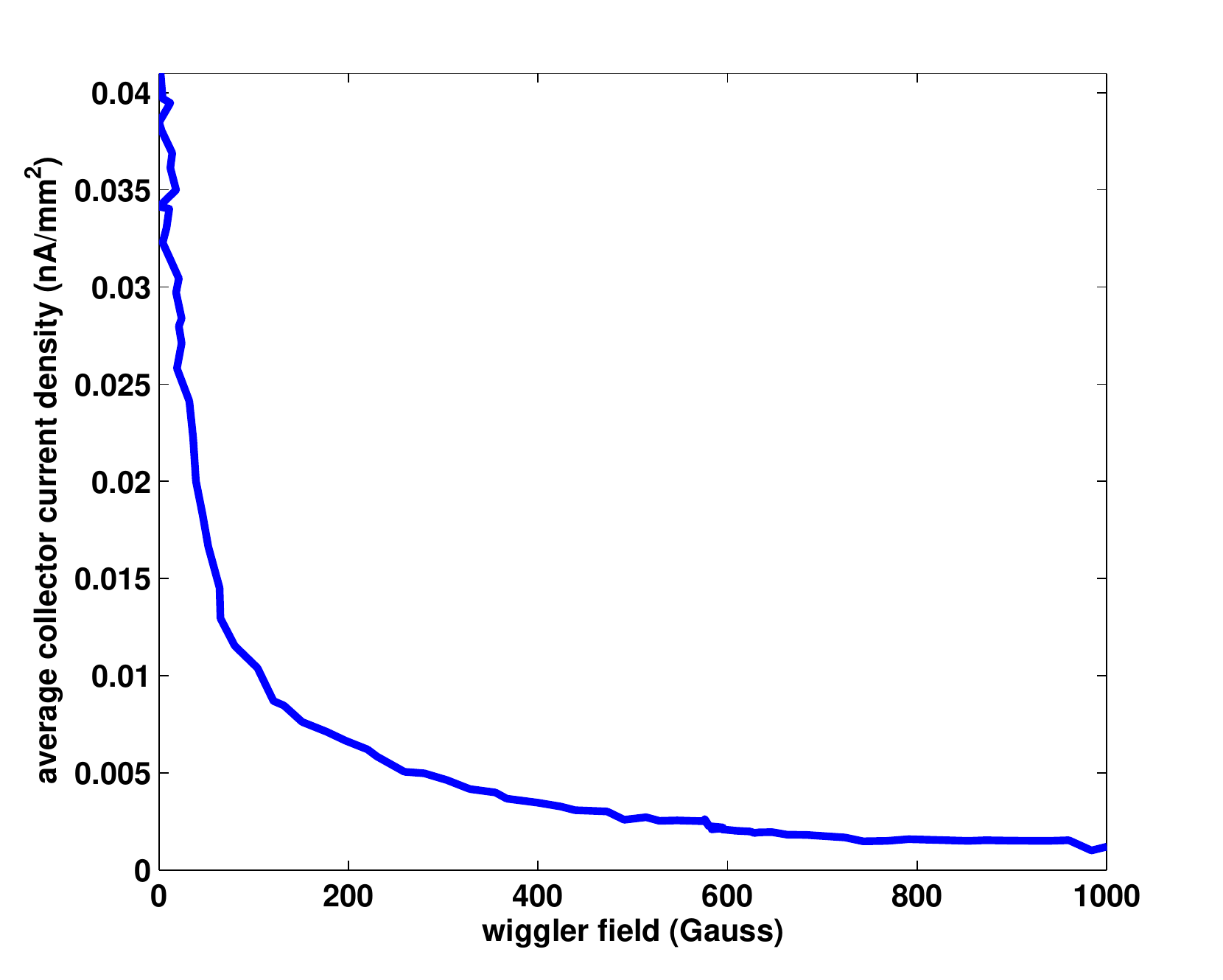} 
   \caption[Wiggler ramp measurement in longitudinal field region]{\label{fig:wig_long} Cu wiggler longitudinal field RFA signal as a function of wiggler strength: 1x45x.75~mA e+, 2.1~GeV, 14~ns.}
\end{figure}

During normal wiggler operation, no signal was ever observed in the RFAs located in the longitudinal field region.  This means that no electrons in the cloud had sufficient transverse energy to cross the longitudinal field lines and reach the vacuum chamber wall.  Fig.~\ref{fig:wig_long} shows the signal in a longitudinal field RFA (in the uncoated copper wiggler), as a function of magnetic field strength.  The signal is effectively gone by 1000 gauss, well below the 1.9~T full field value.  However, simulations have indicated that cloud could be trapped near the beam at these locations~\cite{PRSTAB14:041003}.

\subsubsection{Grooved Chamber Measurement}

\begin{figure}
   \centering
   \includegraphics*[width=0.6\textwidth]{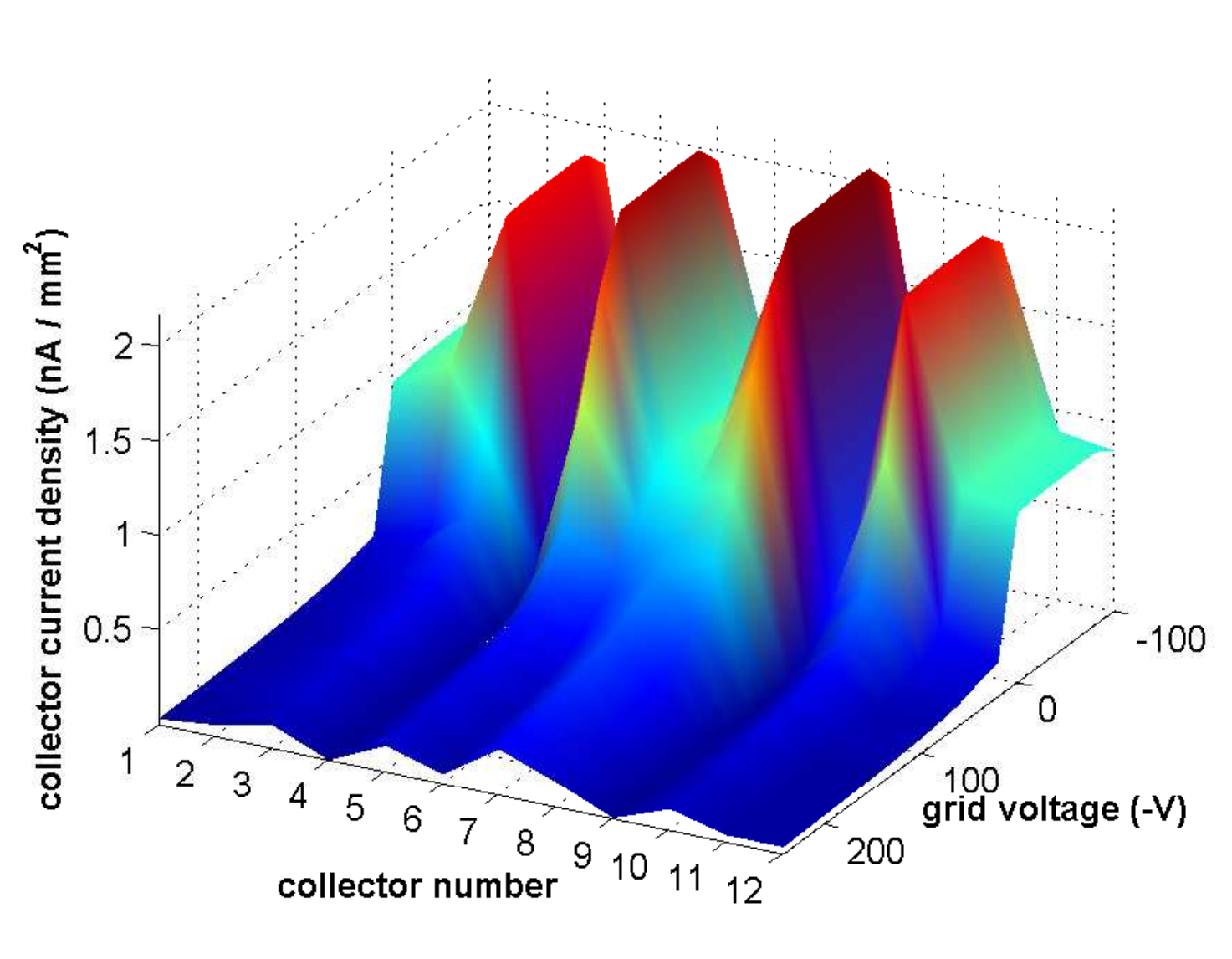} 
   \caption[Grooved Wiggler RFA measurement]{\label{fig:grooved_wig} Grooved wiggler pole center RFA measurement: 1x45x1.25~mA e+, 2.1~GeV, 14~ns.}
\end{figure}

Fig.~\ref{fig:grooved_wig} shows an example voltage scan from the grooved chamber.  The width of the collectors (1.5~mm) is comparable to the distance between peaks of the grooved surface (1~mm), so each collector will sample either two peaks, or two valleys.  Since the valleys are more effective at intercepting electrons than the peaks, we observe alternating high and low signals in the collector currents.  Note that this implies that the effective SEY of the grooves varies significantly across the chamber.

\subsubsection{Clearing Electrode Scan}

\begin{figure}
   \centering
   \includegraphics*[width=0.6\textwidth]{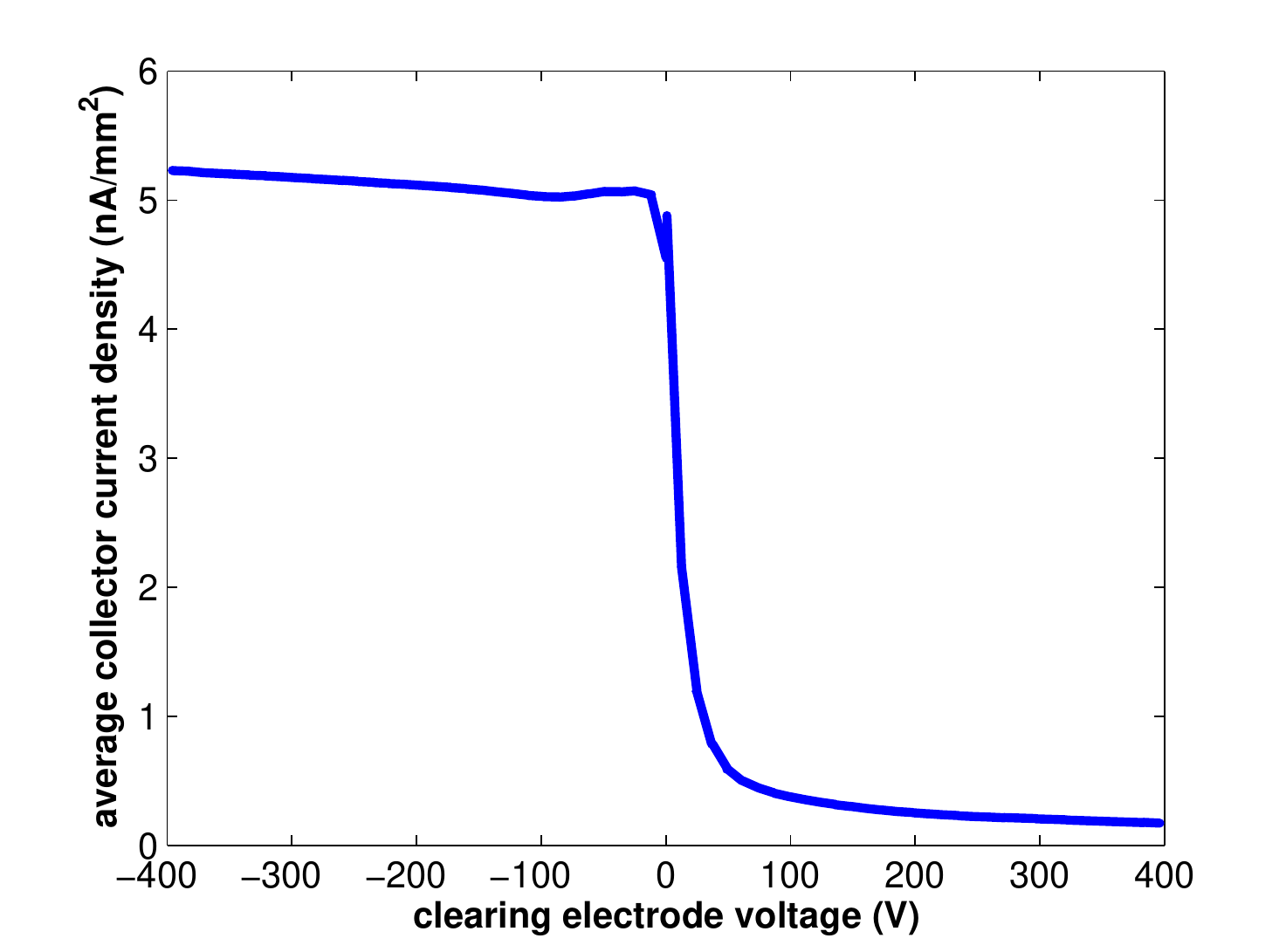} 
   \caption[Clearing electrode scan]{\label{electrode_scan} Clearing electrode scan: 1x45x1~mA e+, 2.1~GeV, 14~ns.}
\end{figure}

Fig.~\ref{electrode_scan} shows the result of varying the clearing electrode voltage on the center pole RFA.  The signal is reduced by an order of magnitude with only +50~V on the electrode, and continues to decrease with higher voltage.  Using a negative voltage actually slightly increases the RFA signal, because the electric field is pushing electrons into the RFA.

\subsubsection{\label{ssec:tramp} Anomalous Enhancement}

\begin{figure}
   \centering
   \begin{tabular}{c}
   \includegraphics*[width=0.6\textwidth]{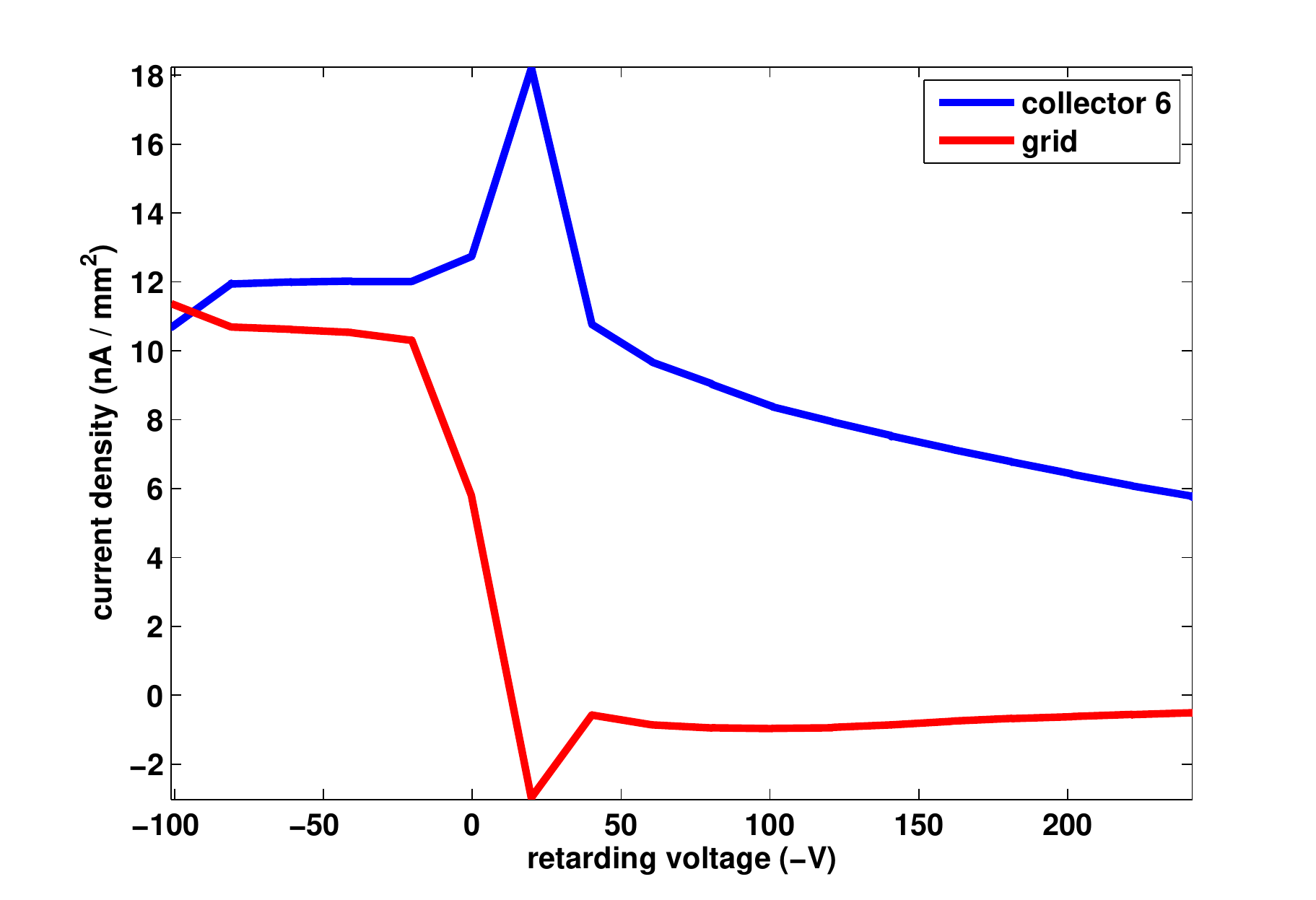}  \\     
   \includegraphics*[width=0.6\textwidth]{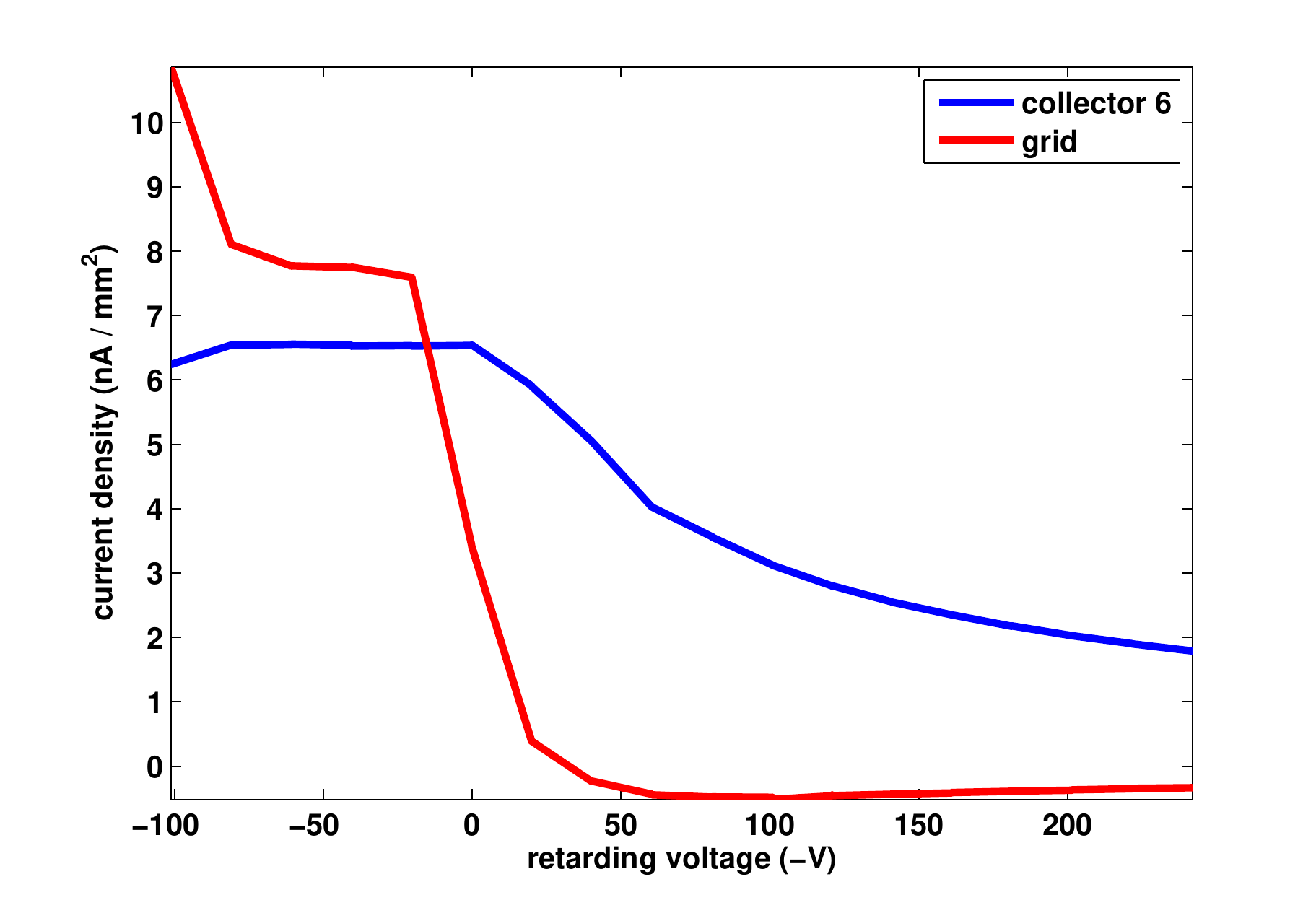}        
   \end{tabular}
   \caption[Resonant enhancement in wiggler data]{\label{fig:tramp_example} Top: Anomalous enhancement in center pole wiggler data.  Bottom: The adjacent wiggler chamber, with high transparency grid, does not show this effect.  Beam conditions are 45 bunches, 1.25~mA/bunch, e+, 2.1~GeV, 14~ns.  Note that there are 12 collectors, so collector 6 is one of the central ones.}
\end{figure}

As described in Section~\ref{ssec:dip_strength}, dipole field strength can have a significant effect on the RFA efficiency.  At very strong fields, unphysical features begin to appear in the data, due to an interaction between the RFA and the cloud it is measuring.  One example of such an interaction can be seen in the top plot of Fig.~\ref{fig:tramp_example}.  It shows a voltage scan done with an RFA in the center pole of a wiggler (in effect a 1.9~T dipole field).  Here one can see a clear enhancement in the signal at low (but nonzero) retarding voltage.  The spike in collector current is accompanied by a corresponding dip in the grid current, suggesting that the grid is the source of the extra collector signal.


More detailed analysis~\cite{CornellU2013:PHD:JCalvey,IPAC10:TUPD022} has shown that the enhancement is caused by a resonance between the bunch spacing and retarding voltage, enabled by the fact that the electron motion is essentially one dimensional.  In short, secondary electrons produced on the grid will be accelerated by the retarding voltage, out the same beam pipe hole the original electron entered.  If the secondary electron is near the center of the chamber during a bunch passage, it will receive a large beam kick, resulting in a high SEY.  Thus an artificial resonance between retarding voltage and bunch spacing is created.

At high beam currents, this effect can become quite significant.  However, it was effectively cured by switching to a high efficiency retarding grid in later wiggler chambers (see Fig.~\ref{fig:tramp_example}, bottom).

\section{Mitigation Comparisons}

Several different electron cloud mitigation techniques have been tested at \cesrta, in dipole, quadrupole, and wiggler environments.  The results shown in this section represent the state of the chambers after beam conditioning (see~\cite{PRL109:064801,NIMA469:1to12,PhysRevSTAB.16.011002} for a discussion of this effect).

\subsection{Dipole Mitigations}

\begin{figure}
      \centering
    \includegraphics*[width=0.6\textwidth]{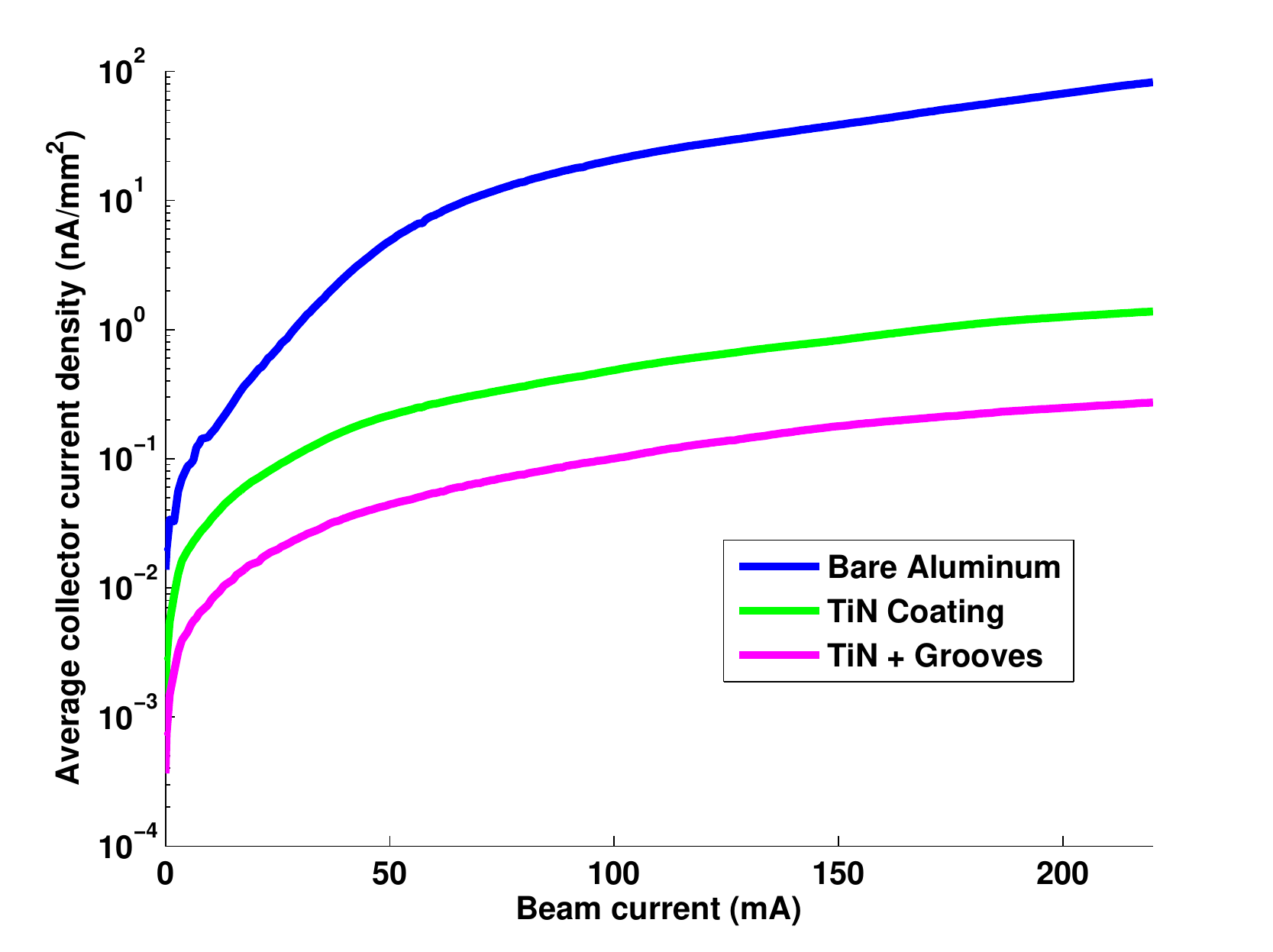}      
   \caption[Dipole RFA mitigation comparison, e+]{\label{fig:dipole_compare}Dipole RFA mitigation comparison, 1x20 e+, 5.3~GeV, 14~ns, 810~gauss.}
\end{figure}

Fig.~\ref{fig:dipole_compare} shows a current scan comparison between three of the chicane RFAs.  We observe a large difference between uncoated and coated chambers.  At high beam current, the TiN coated chamber shows a signal smaller by two orders of magnitude than the bare Al chamber, while the coated and grooved chamber performs better still.

\begin{figure}
      \centering
    \includegraphics*[width=0.6\textwidth]{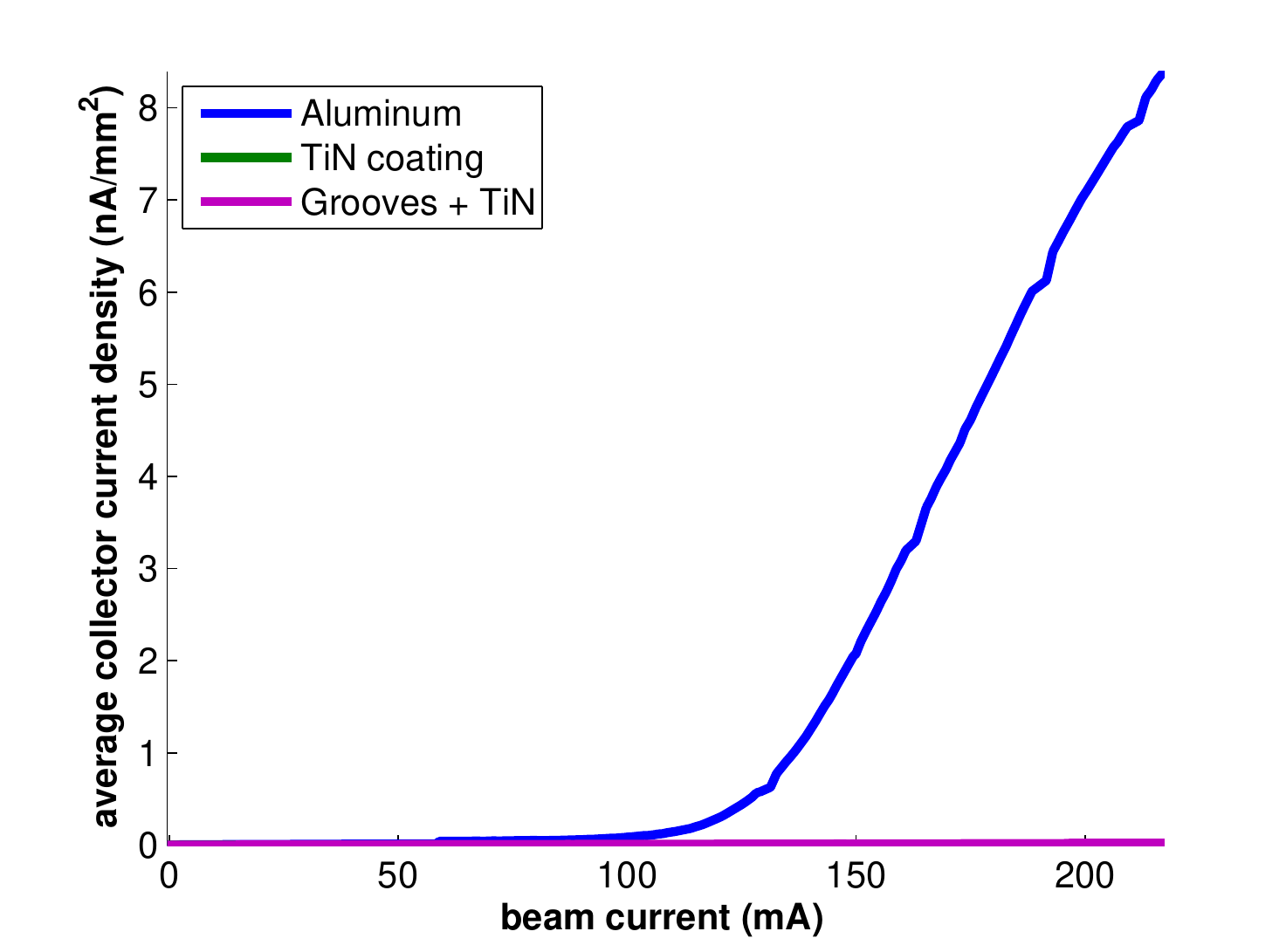} 
   \caption[Dipole RFA mitigation comparison, e-]{\label{fig:dipole_compare_elec}Dipole RFA mitigation comparison, 1x20 e-, 5.3~GeV, 14~ns, 810~gauss.}
\end{figure}

A similar comparison, done with an electron beam, is shown in Fig.~\ref{fig:dipole_compare_elec}.  Here we observe a threshold current, at which the aluminum chamber signal ``turns on," and shows a dramatic increase with current.  This threshold occurs when the kick from the electron beam on low energy cloud particles near the vacuum chamber (which drives them back into the wall) is strong enough to result in significant secondary emission.  The threshold is not observed in any of the chambers with mitigation, where the secondary emission yield is below unity even at high incident energies.


\subsection{Quadrupole Mitigation}

\begin{figure}
   \centering
   \includegraphics*[width=0.6\textwidth]{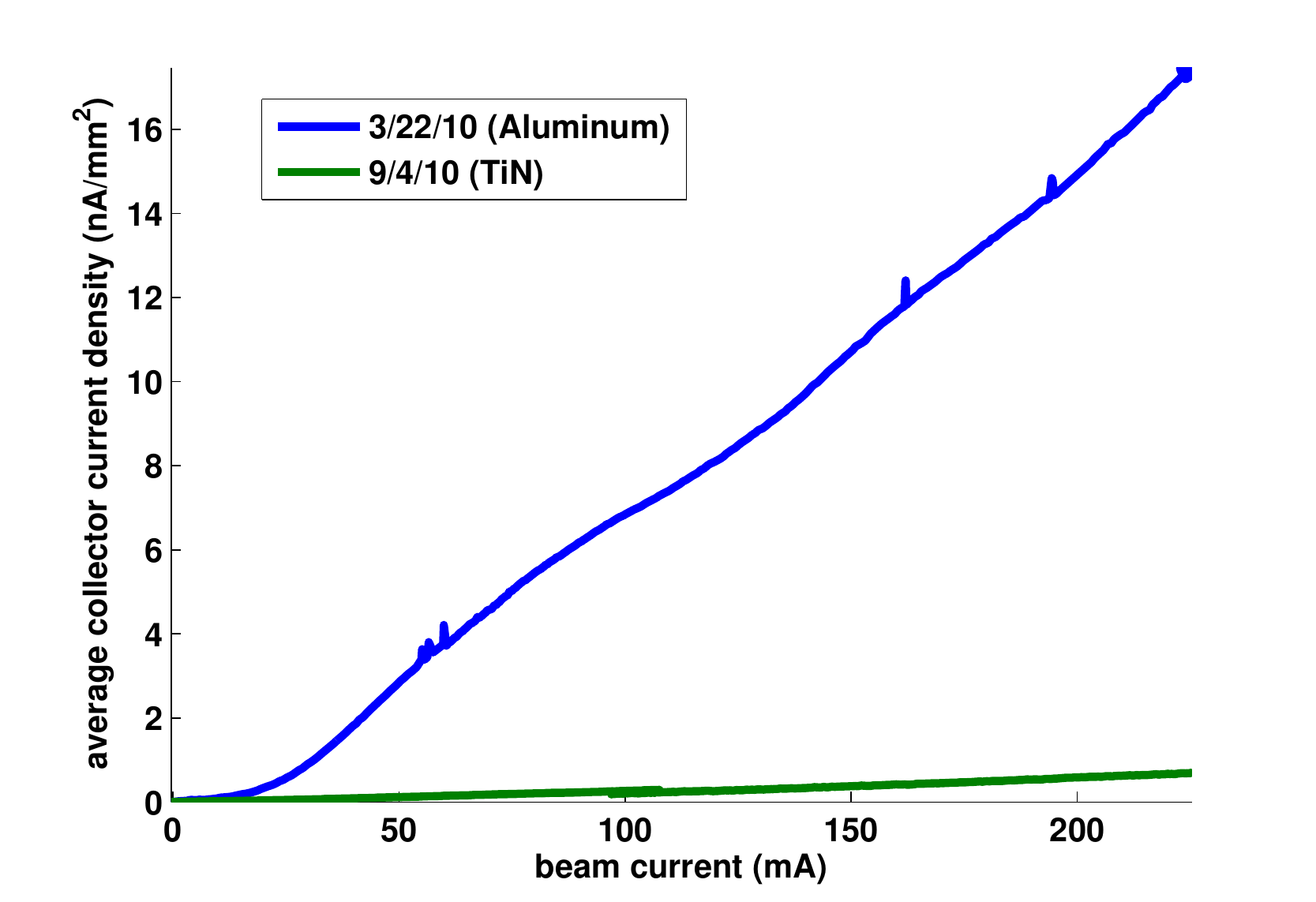}
   \caption[Quadrupole mitigation comparison]{\label{fig:quad_compare} Quadrupole mitigation comparison, 1x20 e+, 5.3~GeV, 14~ns.}
\end{figure}

Fig.~\ref{fig:quad_compare} shows a comparison of a bare aluminum quadrupole chamber with the TiN-coated chamber that replaced it.  The effect of the TiN coating was to reduce the electron cloud signal by well over an order of magnitude.

\subsection{Wiggler Mitigations}

\begin{figure}
    \centering
    \begin{tabular}{c}
    \includegraphics[width=.6\textwidth]{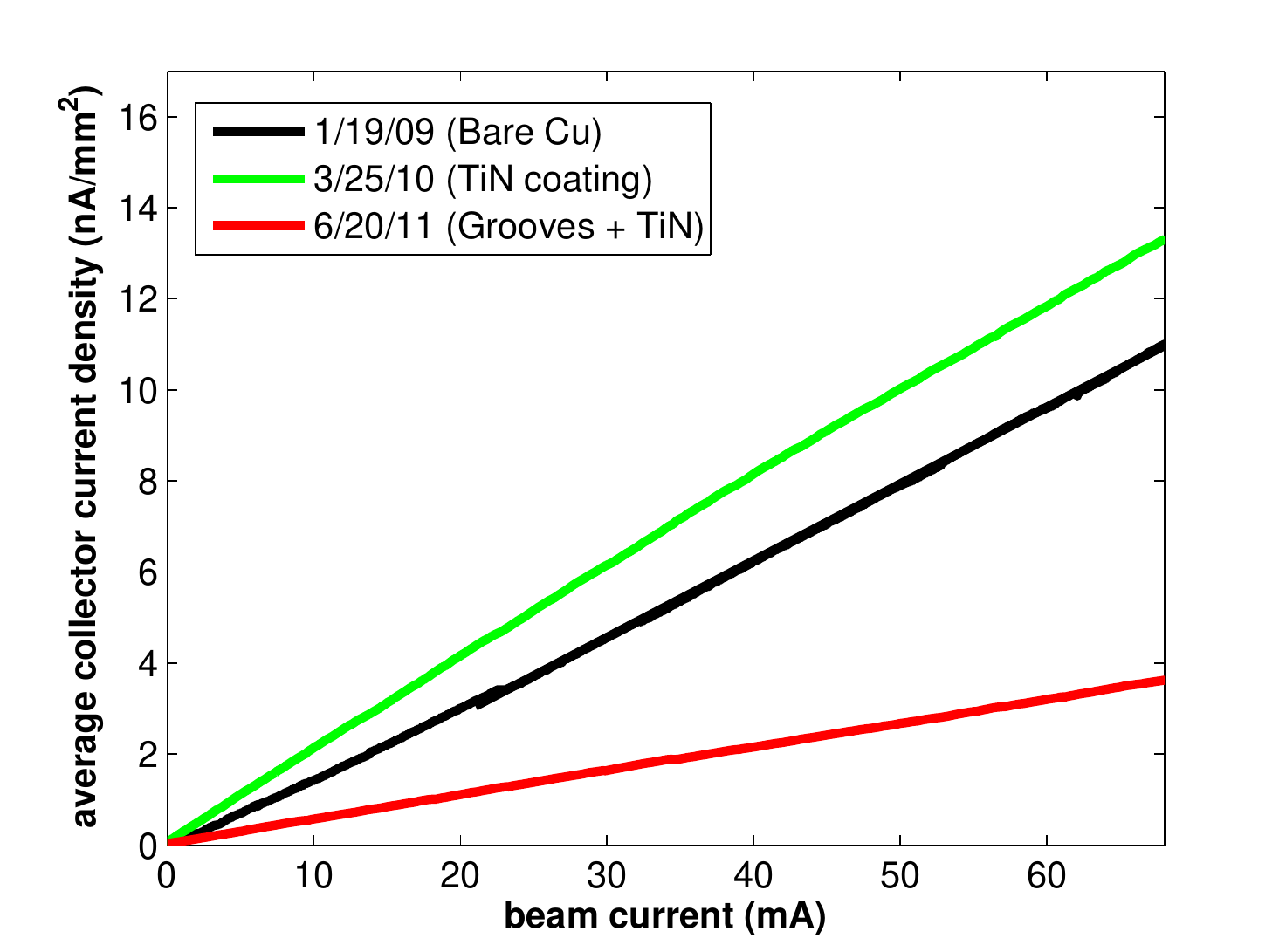} \\
    \includegraphics[width=.6\textwidth]{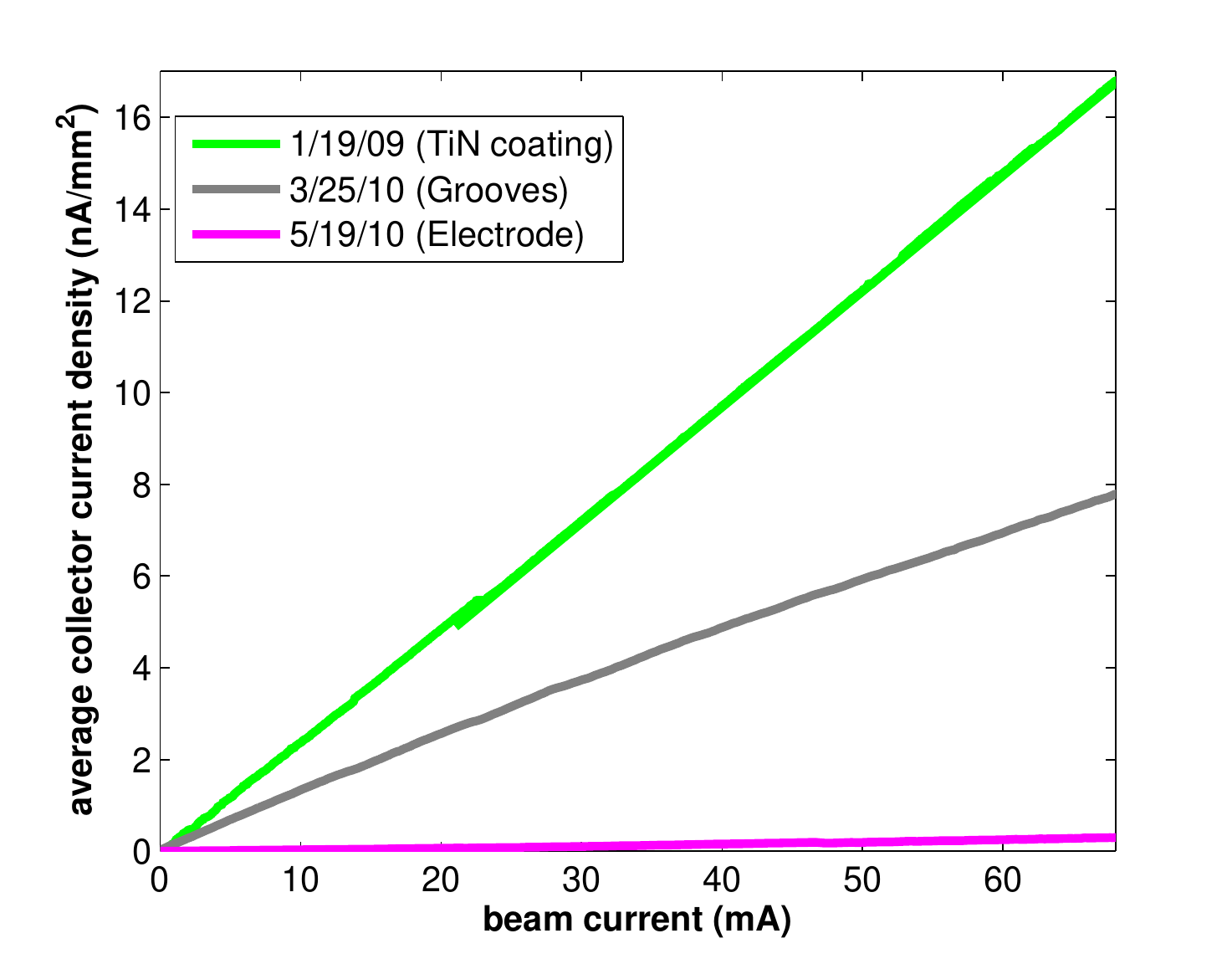} \\
    \end{tabular}
    \caption[Wiggler RFA mitigation comparison]{\label{fig:wig_compare} Wiggler RFA mitigation comparison: 1x45 e+, 2.1~GeV, 14~ns.  Top: 2WA location, bottom: 2WB location.  The 2WB location is further downstream in the wiggler straight, and therefore has a slightly higher photon flux.}
\end{figure}

As described in Section~\ref{ssec:wig_inst}, cycling the location of the different wigglers has allowed us to compare the RFA response with different mitigation techniques at the same longitudinal position in the ring.  Fig.~\ref{fig:wig_compare} compares the average collector current (in the center pole RFA) vs beam current for different mitigation schemes, at both the 2WA and 2WB locations.  These locations have slightly different photon fluxes, but as the TiN coated chamber has been installed in both, it can be used (roughly) as a reference.  Note that TiN coating by itself does not appear to lead to a reduction in the wiggler RFA current relative to bare copper.  This may be due to a reduced sensitivity to SEY at high field, as explained in Section~\ref{ssec:dip_strength}.

Grooves do lead to an improvement, and TiN coated grooves are better still.  The chamber instrumented with a clearing electrode shows the smallest signal by a wide margin, improving on TiN by approximately a factor of 50.  The electrode was set to 400~V for this measurement.  Note that, while TiN is used primarily to reduce secondary emission, both grooves and  electrodes will work equally well on primary and secondary electrons.  This is important for regions that see a large amount of synchrotron radiation, such as a wiggler straight.

\section{Conclusions}

Retarding field analyzers have been installed throughout the CESR ring, in dipole, quadrupole, and wiggler field regions.  Each location presented unique design challenges.  In particular, designing detectors which fit inside the narrow magnet apertures was non-trivial.

A large quantity of electron cloud data has been collected at each location.  To summarize:

\begin{itemize}
    \item Dipole data shows a strong multipacting peak aligned with the beam, where secondary emission is highest.  At high beam currents, this peak bifurcates into two.
    \item The quadrupole data is strongly peaked in the collector aligned with the pole tip, where electrons are guided by the quad field lines.
    \item The center-pole wiggler data shows an anomalous peak at nonzero retarding voltage, caused by secondary emission from the retarding grid.  This problem was solved by using a high efficiency grid.
\end{itemize}

Several different cloud mitigation schemes have been tested.  We observe that:

\begin{itemize}
    \item TiN coating is effective at suppressing cloud growth in a dipole field.  TiN coated grooves show even better suppression.
    \item TiN coating is also effective in a quadrupole.
    \item In a wiggler, the clearing electrode chamber showed the lowest electron cloud signal.
\end{itemize}


In general, we have found that the cloud dynamics, most effective mitigation, and limitations of our detectors all depend on the type and strength of the local magnetic field.  Our results have been incorporated into the proposal for the ILC damping ring~\cite{PRSTAB17:031002}; we expect that they will also be of use in other accelerators.

\section*{Acknowledgements}

The results presented in this paper were made possible by the hard work of the \cesrta~collaboration.  We are especially grateful to X. Liu, B. Clasby, and E. Smith, for their help in constructing the wiggler chambers; and C.R. Strohman, R.E. Meller, S. Santos, and R.M. Schwartz for their help with the RFA measurements.

We would also like to thank Y. Suetsugu at KEK and D. Munson at LBNL, with whom we collaborated in constructing the wiggler chambers; as well as M. Pivi and L. Wang at SLAC, who provided us with the four-dipole chicane.

This research was supported by NSF and DOE Contracts No. PHY-0734867, No. PHY-1002467, No. PHYS-1068662, No. DE-FC02-08ER41538, No. DE-SC0006505, and the Japan/U.S. Cooperation Program.

\appendix

\section{Wiggler Mitigation Details~\label{app:wig}}

\subsection{Grooved Chamber}

\begin{figure}
	\centering
	\includegraphics[width=0.6\textwidth, angle=-90]{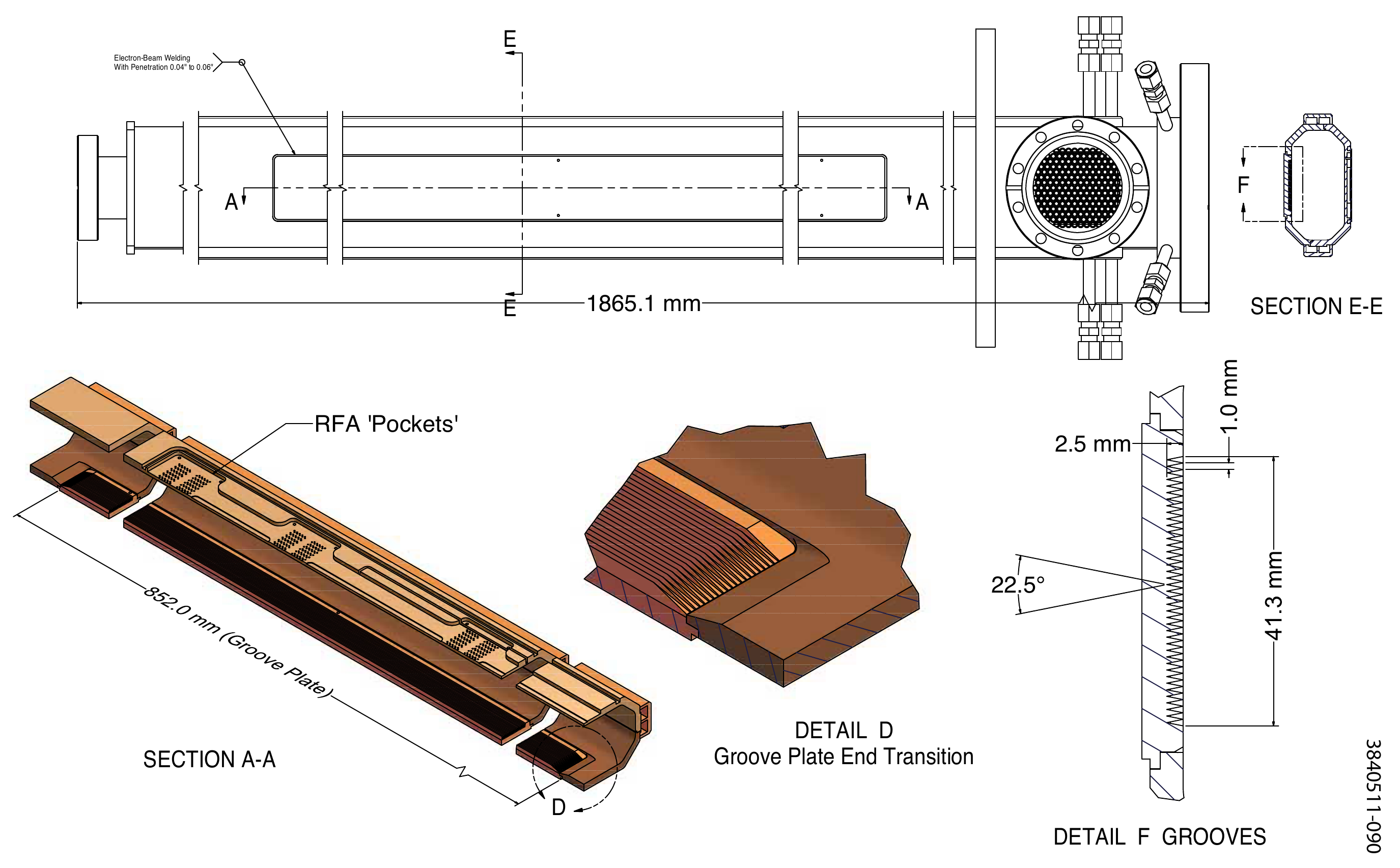}
	\caption{Wiggler RFA beam pipe with a groove plate welded to the bottom beam pipe. \label{fig:cesr_conversion:SCW_RFA_pipe_grooves}}
\end{figure}

Using grooved surfaces to lower effective SEY is a well-known \cite{NIMA571:588to598,Suetsugu2009449,JAP104:104904,PRSTAB7:034401} passive technique to suppress electron cloud growth in a magnetic field.  We studied the effects of grooved chambers by constructing an RFA-equipped wiggler chamber with a grooved insert installed on the bottom surface, as shown in Fig.~\ref{fig:cesr_conversion:SCW_RFA_pipe_grooves}.

The implementation of a grooved surface in a copper vacuum chamber was found to be quite challenging.  First, it was determined (through test machining) that the copper extrusions used for the wiggler beam pipe were fully annealed, and too soft for machining the tiny groove geometry.  Thus the grooves were machined in a separate copper plate of full-hard temper, which was electron-beam welded into the vacuum chamber.  Additionally, it was too costly to machine grooves for the entire length of the chamber.  Thus, for the experimental tests, a grooved plate of sufficient length to span the RFAs was used.


The geometry and dimensions of the grooves are shown in Figure~\ref{fig:cesr_conversion:SCW_RFA_pipe_grooves}, Detail~F.  Cuts at a 30$^\circ$ angle were made on both ends of the groove plate and on the beam pipe wall (shown in Detail~D), which provide a smooth transition from the flat surface to the triangular groove tips in order to minimize higher order mode loss (HOML).


The triangular grooves were made with a milling technique using specially designed cutters.  A small radius of the tips and valleys are desirable for maximum suppression of secondary emission~\cite{NIMA571:588to598}.  We found that we could achieve tip and valley radii of approximately 25~$\mu$m and 60~$\mu$m, which was satisfactory for our purposes.


\begin{figure}
	\centering
\begin{tabular}{cc}
\includegraphics[width=0.45\textwidth]{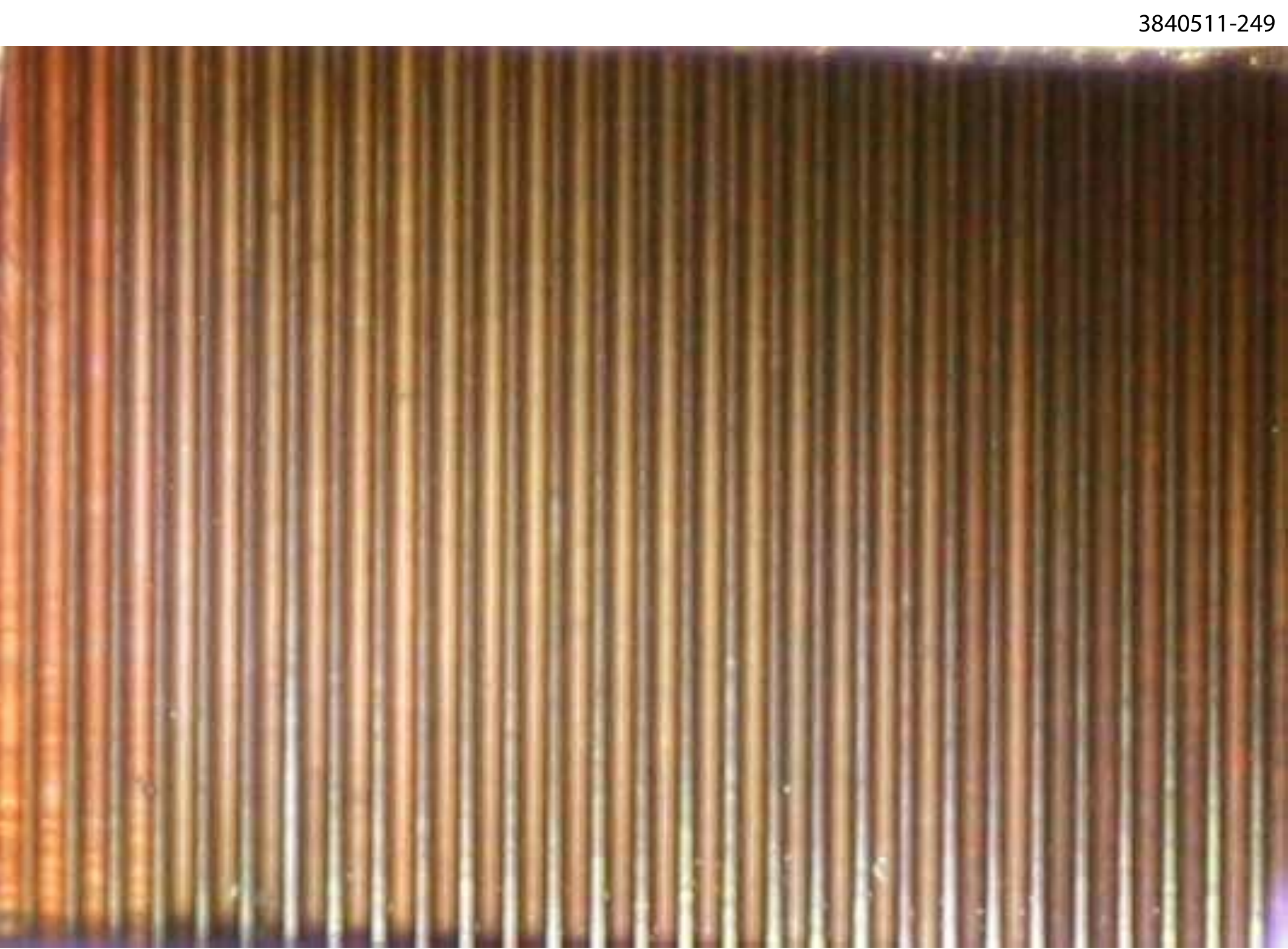} &
\includegraphics[width=0.45\textwidth]{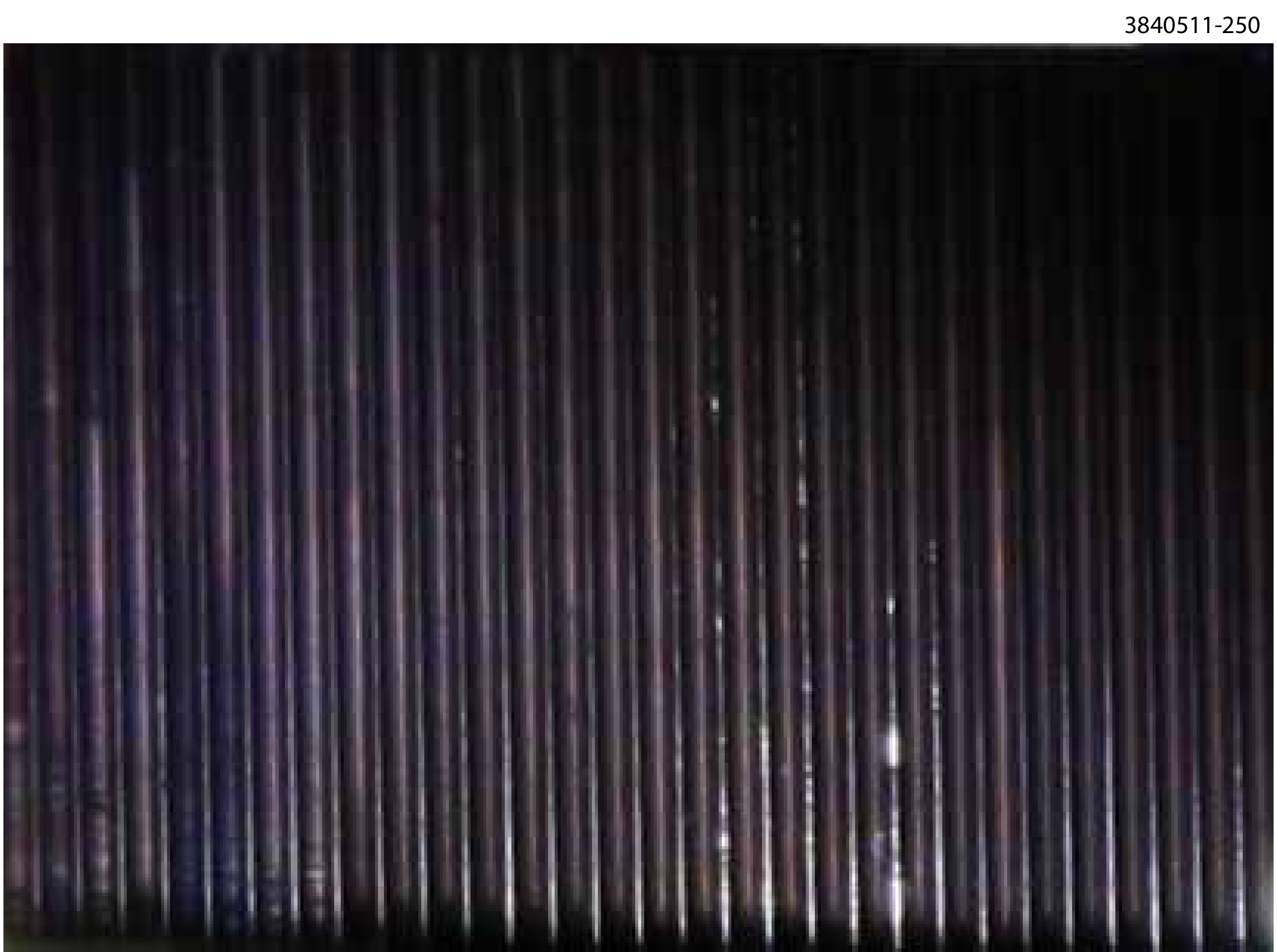}\\
\end{tabular}
	\caption[SCW RFA beam pipe Groove Inspection Photos]{The grooves of the wiggler beam pipe were inspected optically.  Left: Before TiN coating; Right: After TiN coating.  \label{fig:cesr_conversion:SCW_groove_inspection}}
\end{figure}

The wiggler beam pipe assembly, containing the grooves and their associated RFAs, was successfully operated from July~2009 through March~2010, accumulating total beam doses of approximately 940~Amp$\cdot$hr.  After removal, the grooves were inspected for damage (from over-heating), but none was found.  The chamber was then TiN coated, and re-installed in L0.  The TiN deposition was undertaken in two stages with the titanium cathodes shifted transversely to minimize the geometric shadowing of the steep grooves.  The estimated TiN film thickness was 150 to 500~nm, based on the on-line measurements of a QCM (Quartz Crystal Micro-balance) and the off-line measurements of witness coupons.  Close-up optical inspection of the grooves was done after the TiN coating.  Figure~\ref{fig:cesr_conversion:SCW_groove_inspection} shows typical inspection images of a portion of the grooves before and after the TiN coating.  Although the camera did not possess sufficient acuity to resolve the fine features of the sharp tips/troughs, no obvious signs of over-heating or damage to the grooves were visible.  The image also showed relatively uniform TiN coating, with no significant shadowing within the groove troughs.


\subsection{\label{app:electrode} Clearing Electrode}

\begin{figure}
    \centering
    \includegraphics[width=0.75\columnwidth]{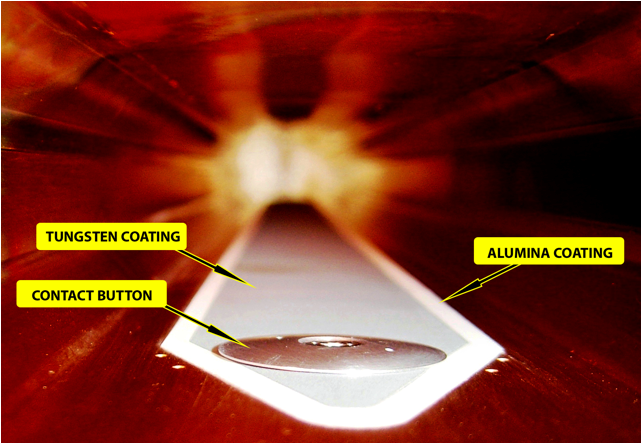}
    \caption[Photo of electrode connection button on the bottom of the wiggler beam pipe.]{\label{fig:SCW_electrode_button} Photograph of the clearing electrode and contact button in the wiggler beam pipe.}
\end{figure}

\begin{figure}
    \centering
    \includegraphics[width=0.6\columnwidth, angle=-90]{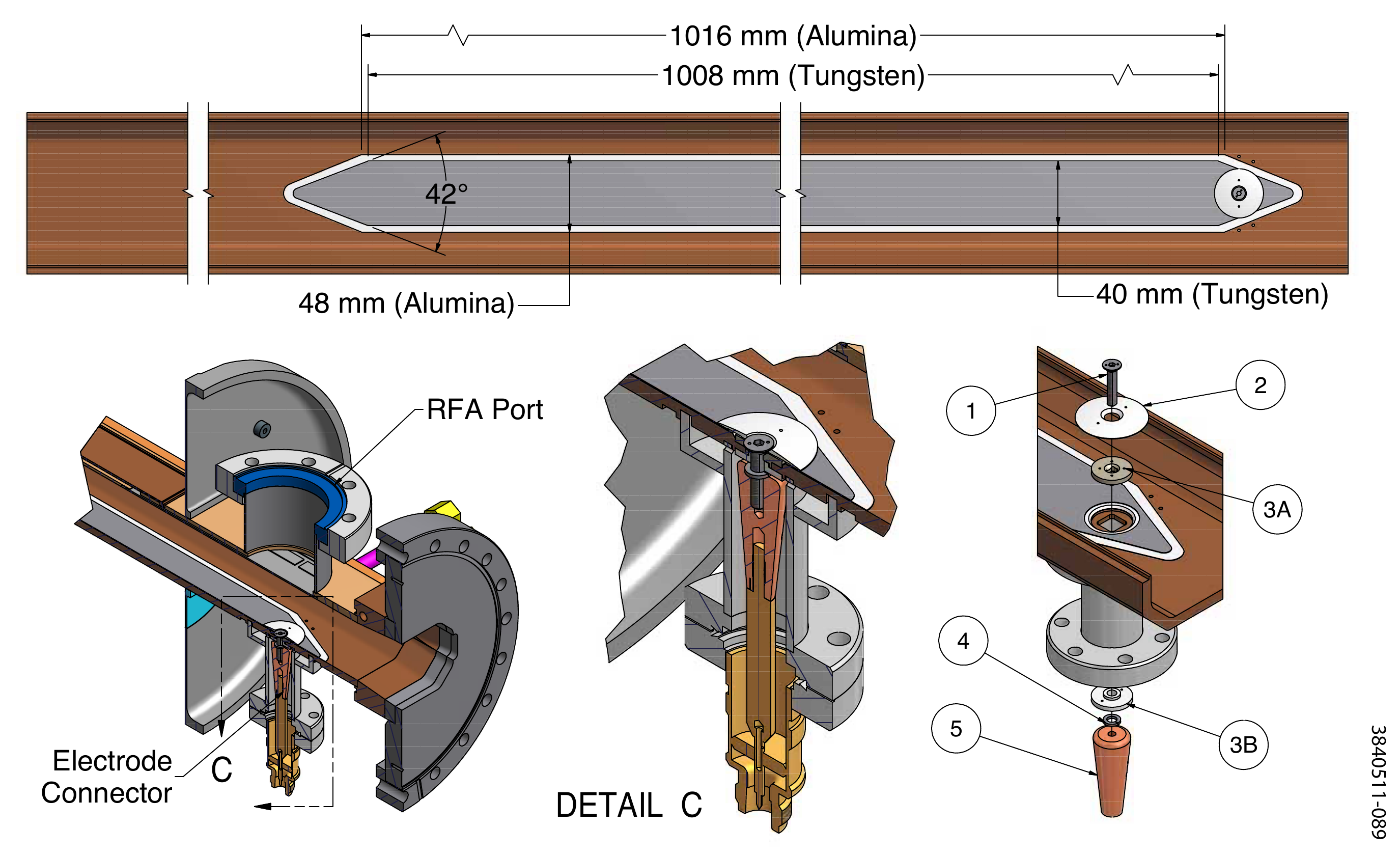}
    \caption[SCW RFA beam pipe with a bottom groove plate with clearing electrode assembly.]{\label{fig:cesr_conversion:SCW_RFA_pipe_electrode}
	Wiggler RFA beam pipe with clearing electrode assembly.  Top: sectional view of the clearing electrode as seen from above. Bottom left: sectional view  of the clearing electrode and the HV input port. Bottom middle: detail of HV port. Bottom right: exploded view of HV port;  electrical contact is made with a vented screw (1), an aluminum alloy button (2), insulation spacers made of PEEK (3A, 3B),  a stainless steel spring washer (4), and a tapered copper pin (5).}
\end{figure}

In April 2010, a clearing electrode, shown in Fig.~\ref{fig:SCW_electrode_button}, was installed on the bottom of an RFA-equipped wiggler chamber.  The structure of the electrode equipped chamber is diagrammed in Figure~\ref{fig:cesr_conversion:SCW_RFA_pipe_electrode}.



The clearing electrode forms a thin stripline, consisting of an insulating alumina layer ($\sim 0.2$~mm thick) deposited on the copper beam chamber, and a conducting layer of tungsten ($\sim 0.1$~mm thick, deposited on the alumina) as the electrode.  The area of the alumina ceramic layer was larger than that of the electrode, so that the required DC voltage could be applied to the electrode in vacuum.  These two layers were deposited via a thermal spray technique developed at KEK~\cite{NIMA598:372to378}, and were tightly bonded to the copper chamber.  Tungsten was chosen as the electrode material owing to its good thermal and electrical properties, in particular its small thermal expansion rate.  The insulation resistance between the electrode and the copper chamber was about 5~M$\Omega$ (in dry air), and the electrode is capable of withholding DC voltages above 1~kV.



The clearing electrode is designed to have a width of 40~mm and is placed on the lower wall of the vacuum chamber, extending for 1.09~m along the beamline, in order to intercept electrons impacting the bottom of the vacuum chamber.  To minimize higher order mode loss (HOML) induced by the electrode, the ends of the electrode are tapered at a 42$^\circ$ angle down to a 3~mm radius at its tip.
The high voltage (HV) connection is made by a coaxial line coming through a port located underneath one of the tapered ends of the electrode.  The electrical connection to the electrode is made via a convex button washer (having a 26~mm diameter) on the top of the electrode, as shown in Figure~\ref{fig:cesr_conversion:SCW_RFA_pipe_electrode}.  The connection was designed to make the inner surface (visible to the beam) as smooth as possible, while keeping a secure electrical contact.  Although the effects of the low profile of the discontinuity, the tapered ends of the electrode and the hidden HV connection have not been calculated in detail, they are expected to produce a HOML parameter of less than a few times 0.001~V/pC for the 7 to 10~mm bunch length of CESR.  Any heating from the wall currents flowing on the surface of the electrode is easily handled by conduction to the beam pipe through the thin alumina dielectric layer, which has good thermal conductivity.  Thermocouples mounted on the bottom beam pipe near the electrode assembly have detected no increase in heating due to HOML during operations.  A detailed account of the impedance characteristics of the clearing electrode is given in~\cite{CLNS:12:2084}.

The clearing electrode was operated from April to December 2010, accumulating a beam dose in excess of 1000 Amp$\cdot$hr.  In January 2011, a visual inspection of the electrode and the electrical contacts was done.  Both were found to be in excellent condition with no sign of arcing or overheating.  The electrode chamber has remained in CESR for three years since then, with no noticeable degradation of performance.

\bibliographystyle{medium}
\bibliography{rfa_field_nim}

\begin{thebibliography}{10}
\newcommand{\enquote}[1]{``#1''}
\providecommand{\url}[1]{\texttt{#1}}
\providecommand{\urlprefix}{URL }
\providecommand{\eprint}[2][]{\url{#2}}

\bibitem{ECLOUD12:Tue1815}
M.~A. Furman, \enquote{Electron Cloud Effects in Accelerators,} in
  \emph{Proceedings of ECLOUD 2012: Joint INFN-CERN-EuCARD-AccNet Workshop on
  Electron-Cloud Effects, La Biodola, Elba, Italy}, R.~Cimino, G.~Rumolo \&
  F.~Zimmermann, Eds., CERN, Geneva (2013), CERN-2013-002, p. 1--8.

\bibitem{PhysRevSTAB.7.124801}
F.~Zimmermann, \enquote{Review of single bunch instabilities driven by an
  electron cloud,} \emph{Phys. Rev. ST Accel. Beams} \textbf{7}, 124801 (Dec.
  2004).

\bibitem{ILCREP2007:001}
G.~Aarons \emph{et~al.}, \enquote{International Linear Collider Reference
  Design Report,} Tech. Rep. ILC-REPORT-2007-001, International Linear
  Collider, Batavia, IL/Tsukuba, Japan/Hamburg, Germany (Aug. 2007).

\bibitem{CLNS:12:2084}
\enquote{The CESR Test Accelerator Electron Cloud Research Program: Phase I
  Report,} Tech. Rep. CLNS-12-2084, LEPP, Cornell University, Ithaca, NY (Jan.
  2013).

\bibitem{NIMA760:86to97}
J.~R. Calvey \emph{et~al.}, \enquote{Comparison of Electron Cloud Mitigating
  Coatings Using Retarding Field Analyzers,} \emph{Nucl. Instrum. Methods Phys.
  Res.} \textbf{A760}, p. 86--97 (Oct. 2014).

\bibitem{PRSTAB17:061001}
J.~R. Calvey \emph{et~al.}, \enquote{Measurement and Modeling of Electron Cloud
  in a Field Free Environment Using Retarding Field Analyzers,} \emph{Phys.
  Rev. ST Accel. Beams} \textbf{17}, 061001 (Jun. 2014).

\bibitem{NIMA749:42to46}
J.~A. Crittenden \emph{et~al.}, \enquote{Shielded Button Electrodes for
  Time-Resolved Measurements of Electron Cloud Buildup,} \emph{Nucl. Instrum.
  Methods Phys. Res.} \textbf{A749}, p. 42--46 (Jun. 2014).

\bibitem{ARXIV:1407.0772}
W.~Hartung \emph{et~al.}, \enquote{Instrumentation and Methods for In-Situ
  Measurements of the Secondary Electron Yield in an Accelerator Environment,}
  Tech. Rep. arXiv:1407.0772, Cornell University Library, Ithaca, New York
  (Jul. 2014), submitted to Nucl. Instrum. Methods. Phys. Res. A.

\bibitem{NIMA754:28to35}
J.~P. Sikora \emph{et~al.}, \enquote{Electron Cloud Density Measurements in
  Accelerator Beam-Pipe Using Resonant Microwave Excitation,} \emph{Nucl.
  Instrum. Methods Phys. Res.} \textbf{A754}, p. 28--35 (Aug. 2014).

\bibitem{NIMA453:507to513}
R.~A. Rosenberg \& K.~C. Harkay, \enquote{A Rudimentary Electron Energy
  Analyzer for Accelerator Diagnostics,} \emph{Nucl. Instrum. Methods Phys.
  Res.} \textbf{A453}, p. 507--513 (Oct. 2000).

\bibitem{NIMA621:33to38}
M.~T.~F. Pivi \emph{et~al.}, \enquote{Observation of Magnetic Resonances in
  Electron Clouds in a Positron Storage Ring,} \emph{Nucl. Instrum. Methods
  Phys. Res.} \textbf{A621}, p. 33--38 (Sep. 2010).

\bibitem{NIMA598:372to378}
Y.~Suetsugu \emph{et~al.}, \enquote{Demonstration of Electron Clearing Effect
  by Means of a Clearing Electrode in High-Intensity Positron Ring,}
  \emph{Nucl. Instrum. Methods Phys. Res.} \textbf{A598}, p. 372--378 (Jan.
  2009).

\bibitem{Suetsugu2009449}
Y.~Suetsugu \emph{et~al.}, \enquote{Continuing study on electron-cloud clearing
  techniques in high-intensity positron ring: Mitigation by using groove
  surface in vertical magnetic field,} \emph{Nucl. Instrum. Methods Phys. Res.}
  \textbf{604}, p. 449 -- 456 (2009).

\bibitem{PhysRevLett.110.124801}
D.~Alesini \emph{et~al.}, \enquote{DA$\Phi${}NE Operation with
  Electron-Cloud-Clearing Electrodes,} \emph{Phys. Rev. Lett.} \textbf{110},
  124801 (Mar 2013).

\bibitem{PhysRevSTAB.11.094401}
E.~Mahner, T.~Kroyer \& F.~Caspers, \enquote{Electron cloud detection and
  characterization in the CERN Proton Synchrotron,} \emph{Phys. Rev. ST Accel.
  Beams} \textbf{11}, 094401 (Sep 2008).

\bibitem{ECLOUD02:17to28}
J.~M. Jimenez \emph{et~al.}, \enquote{Electron Cloud with LHC-Type Beams in the
  SPS: A Review of Three Years of Measurements,} in \emph{Mini Workshop on
  Electron Cloud Simulations for Proton and Positron Beams---ECLOUD'02, Geneva,
  Switzerland}, G.~Rumolo \& F.~Zimmermann, Eds., CERN, Geneva, Switzerland
  (2002), CERN-2002-001, p. 17--28.

\bibitem{PhysRevSTAB.11.010101}
R.~J. Macek \emph{et~al.}, \enquote{Electron cloud generation and trapping in a
  quadrupole magnet at the Los Alamos proton storage ring,} \emph{Phys. Rev. ST
  Accel. Beams} \textbf{11}, 010101 (Jan 2008).

\bibitem{PhysRevLett.97.054801}
M.~Kireeff~Covo \emph{et~al.}, \enquote{Absolute Measurement of Electron-Cloud
  Density in a Positively Charged Particle Beam,} \emph{Phys. Rev. Lett.}
  \textbf{97}, 054801 (Jul 2006).

\bibitem{PRSTAB14:041003}
C.~M. Celata, \enquote{Electron Cloud Dynamics in the Cornell Electron Storage
  Ring Test Accelerator Wiggler,} \emph{Phys. Rev. ST Accel. Beams}
  \textbf{14}, 041003 (Apr. 2011).

\bibitem{NIMA551:187to199}
F.~Le~Pimpec \emph{et~al.}, \enquote{Properties of TiN and TiZrV Thin Film as a
  Remedy Against Electron Cloud,} \emph{Nucl. Instrum. Methods Phys. Res.}
  \textbf{A551}, p. 187--199 (Jul. 2005).

\bibitem{NIMA578:470to479}
Y.~Suetsugu \emph{et~al.}, \enquote{Recent Studies on Photoelectron and
  Secondary Electron Yields of TiN and NEG Coatings Using the KEKB Positron
  Ring,} \emph{Nucl. Instrum. Methods Phys. Res.} \textbf{A578}, p. 470--479
  (Aug. 2007).

\bibitem{NIMA556:399to409}
Y.~Suetsugu \emph{et~al.}, \enquote{Continuing Study on the Photoelectron and
  Secondary Electron Yield of TiN Coating and NEG (Ti-Zr-V) Coating Under
  Intense Photon Irradiation at the KEKB Positron Ring,} \emph{Nucl. Instrum.
  Methods Phys. Res.} \textbf{A556}, p. 399--409 (Jan. 2006).

\bibitem{NIMA564:44to50}
F.~Le~Pimpec \emph{et~al.}, \enquote{The Effect of Gas Ion Bombardment on the
  Secondary Electron Yield of TiN, TiCN and TiZrV Coatings for Suppressing
  Collective Electron Effects in Storage Rings,} \emph{Nucl. Instrum. Methods
  Phys. Res.} \textbf{A564}, p. 44--50 (Aug. 2006).

\bibitem{NIMA571:588to598}
L.~Wang, T.~O. Raubenheimer \& G.~Stupakov, \enquote{Suppression of Secondary
  Emission in a Magnetic Field Using Triangular and Rectangular Surfaces,}
  \emph{Nucl. Instrum. Methods Phys. Res.} \textbf{A571}, p. 588--598 (Feb.
  2007).

\bibitem{JAP104:104904}
M.~Pivi \emph{et~al.}, \enquote{Sharp Reduction of the Secondary Electron
  Emission Yield from Grooved Surfaces,} \emph{J. Appl. Phys.} \textbf{104},
  104904 (Nov. 2008).

\bibitem{PRSTAB7:034401}
L.~F. Wang \emph{et~al.}, \enquote{Mechanism of Electron Cloud Clearing in the
  Accumulator Ring of the Spallation Neutron Source,} \emph{Phys. Rev. ST
  Accel. Beams} \textbf{7}, 034401 (Mar. 2004).

\bibitem{PAC07:Kryoer}
T.~Kroyer \emph{et~al.}, \enquote{A New Type of Distributed Enamel Based
  Clearing Electrode,} in \emph{Proceedings of the 2007 Particle Accelerator
  Conference, Albuquerque, NM} (June 2007), p. 2000--2002.

\bibitem{IPAC12:WEPPR088}
J.~R. Calvey \emph{et~al.}, \enquote{Modeling and Simulation of Retarding Field
  Analyzers at CESRTA,} in \emph{Proceedings of the 2012 International Particle
  Accelerator Conference, New Orleans, LA} (2012), p. 3138--3140.

\bibitem{IPAC11:MOPS083}
J.~R. Calvey \emph{et~al.}, \enquote{Update on Electron Cloud Mitigation
  Studies at Cesr-TA,} in \emph{Proceedings of the 2011 International Particle
  Accelerator Conference, San {Sebasti\'{a}n}, Spain} (2011), p. 796--798.

\bibitem{NAPAC13:FROAA5}
J.~P. Sikora \emph{et~al.}, \enquote{Electron Cloud Measurements Using a
  Shielded Pickup in a Quadrupole at CesrTA,} in \emph{NA-PAC 2013: Proceedings
  of the 2013 North American Particle Accelerator Conference, Pasadena, CA},
  T.~Satogata, C.~Petit-Jean-Genaz \& V.~Schaa, Eds. (2013), p. 1437--1439.

\bibitem{CornellU2013:PHD:JCalvey}
J.~Calvey, \emph{Studies of Electron Cloud Growth and Mitigation at Cesr-TA},
  Ph.D. thesis, Cornell University, Ithaca, New York (Aug. 2013).

\bibitem{IPAC10:TUPD022}
J.~R. Calvey \emph{et~al.}, \enquote{CesrTA Retarding Field Analyzer Modeling
  Results,} in \emph{Proceedings of the 2010 International Particle Accelerator
  Conference, Kyoto, Japan} (2010), p. 1970--1972.

\bibitem{PRL109:064801}
R.~Cimino \emph{et~al.}, \enquote{Nature of the Decrease of the
  Secondary-Electron Yield by Electron Bombardment and its Energy Dependence,}
  \emph{Phys. Rev. Lett.} \textbf{109}, 064801 (Aug. 2012).

\bibitem{NIMA469:1to12}
R.~E. Kirby \& F.~K. King, \enquote{Secondary Electron Emission Yields from
  PEP-II Accelerator Materials,} \emph{Nucl. Instrum. Methods Phys. Res.}
  \textbf{A469}, p. 1--12 (Aug. 2001).

\bibitem{PhysRevSTAB.16.011002}
R.~Larciprete \emph{et~al.}, \enquote{Secondary electron yield of Cu technical
  surfaces: Dependence on electron irradiation,} \emph{Phys. Rev. Lett.}
  \textbf{16}, 011002 (Jan. 2013).

\bibitem{PRSTAB17:031002}
J.~A. Crittenden \emph{et~al.}, \enquote{Investigation into Electron Cloud
  Effects in the International Linear Collider Positron Damping Ring,}
  \emph{Phys. Rev. ST Accel. Beams} \textbf{17}, 031002 (Mar. 2014).

\end{thebibliography}

\end{document}